\RequirePackage{fix-cm}
\documentclass[twocolumn, epjc3, comma, sort&compress, natbib]{svjour3}
\bibpunct{[}{]}{,}{n}{}{,} 

\journalname{Eur. Phys. J. C}

\usepackage{latexsym}
\usepackage{amsmath}
\usepackage{amssymb}
\usepackage{amsfonts}
\usepackage{CJKutf8}

\usepackage[mathscr,scaled=1.15]{urwchancal}
\DeclareFontFamily{OT1}{pzc}{}
\DeclareFontShape{OT1}{pzc}{m}{it}%
{<-> s * [1.15] pzcmi7t}{}
\DeclareMathAlphabet{\mathpzc}{OT1}{pzc}{m}{it}

\usepackage{color}

\usepackage{supertabular}
\usepackage{placeins}
\usepackage{epsfig}
\usepackage{graphicx}

\definecolor{purple}{rgb}{0.5,0,0.5}
\definecolor{blue}{rgb}{0.0,0,0.9}
\definecolor{prdblue}{rgb}{0.133,0.118,0.498}
\usepackage[colorlinks=true, pdfstartview=FitV, linkcolor=prdblue, citecolor= prdblue, urlcolor=prdblue]{hyperref}

\begin{document}
\begin{CJK}{UTF8}{song}

\title{$\,$\\[-6ex]\hspace*{\fill}{\normalsize{\sf\emph{Preprint no}.\ NJU-INP 081/23}}\\[1ex]
Pion and kaon electromagnetic and gravitational form factors}

\author{Y-Z.~Xu\thanksref{Huelva,UPO}
    $^{\href{https://orcid.org/0000-0003-1623-3004}{\textcolor[rgb]{0.00,1.00,0.00}{\sf ID}},}$
    \and
        M.~Ding\thanksref{HZDR}
       $^{\href{https://orcid.org/0000-0002-3690-1690}{\textcolor[rgb]{0.00,1.00,0.00}{\sf ID}},}$
    \and
        K.~Raya\thanksref{Huelva}%
    $^{\href{https://orcid.org/0000-0001-8225-5821}{\textcolor[rgb]{0.00,1.00,0.00}{\sf ID}},}$
    \and
    C.\,D.~Roberts\thanksref{NJU,INP}%
    $^{\href{https://orcid.org/0000-0002-2937-1361}{\textcolor[rgb]{0.00,1.00,0.00}{\sf ID}},}$
    \and
        J.~Rodr\'iguez-Quintero\thanksref{Huelva}%
       $^{\href{https://orcid.org/0000-0002-1651-5717}{\textcolor[rgb]{0.00,1.00,0.00}{\sf ID}},}$
    \and
        S.\,M.~Schmidt\thanksref{HZDR,RWTH}%
       $^{\href{https://orcid.org/0000-0002-8947-1532}{\textcolor[rgb]{0.00,1.00,0.00}{\sf ID}},}$
}

\authorrunning{Y.-Z. Xu \emph{et al}.} 

\institute{Dpto. Ciencias Integradas, Centro de Estudios Avanzados en Fis., Mat. y Comp., Fac. Ciencias Experimentales, \\ \hspace*{0.5em}Universidad de Huelva, Huelva 21071, Spain \label{Huelva}
    \and
    Dpto. Sistemas F\'isicos, Qu\'imicos y Naturales, Univ.\ Pablo de Olavide, E-41013 Sevilla, Spain \label{UPO}
            \and
            Helmholtz-Zentrum Dresden-Rossendorf, Bautzner Landstra{\ss}e 400, D-01328 Dresden, Germany \label{HZDR}
            \and
            School of Physics, Nanjing University, Nanjing, Jiangsu 210093, China \label{NJU}
           \and
           Institute for Nonperturbative Physics, Nanjing University, Nanjing, Jiangsu 210093, China \label{INP}
           \and
           RWTH Aachen University, III. Physikalisches Institut B, D-52074 Aachen, Germany \label{RWTH}
\\[1ex]
Email:
\href{mailto:m.ding@hzdr.de}{m.ding@hzdr.de} (MD);
\href{mailto:cdroberts@nju.edu.cn}{cdroberts@nju.edu.cn} (C. D. Roberts);
\href{mailto:jose.rodriguez@dfaie.uhu.es}{jose.rodriguez@dfaie.uhu.es} (J. Rodr\'iguez-Quintero)
            }

\date{2023 November 25}

\maketitle

\end{CJK}

\begin{abstract}
A unified set of predictions for pion and kaon elastic electromagnetic and gravitational form factors is obtained using a symmetry-preserving truncation of each relevant quantum field equation.  A key part of the study is a description of salient aspects of the dressed graviton + quark vertices.  The calculations reveal that each meson's mass radius is smaller than its charge radius, matching available empirical inferences; and meson core pressures are commensurate with those in neutron stars.  The analysis described herein paves the way for a direct calculation of nucleon gravitational form factors. 
\end{abstract}


\section{Introduction}

A new era is heralded by construction and planning of high-luminosity, high-energy facilities \cite{Adams:2018pwt, Brodsky:2020vco, Chen:2020ijn, Anderle:2021wcy, Arrington:2021biu, Quintans:2022utc, Wang:2022xad, Chang:2022pcb, Carman:2023zke, Accardi:2023chb}.  Following a one-hundred year focus on the proton, such facilities will enable exploration of the structure of many other states produced by the strong interaction Hamiltonian.  Principal amongst them are pions ($\pi$) and kaons ($K$), \emph{viz}.\ Nature's most fundamental (would-be) Nambu-Gold\-stone bosons.  These states are special because, absent Higgs boson (HB) couplings into quantum chromodynamics (QCD), they are all massless and identical.  The Higgs inserts quark current-masses into the QCD Hamiltonian; yet at realistic current-mass values, $\pi$- and $K$ mesons remain unnaturally light, with masses that seemingly belie their categorisation as hadrons, \emph{e.g}., the $\pi$ mass is similar to that of the $\mu$-lepton, despite pions being strong-interaction bound-states.

Contemporary theory argues that the natural mass scale for strong interactions, \emph{i.e}., the proton mass, $m_p \approx 1\,$GeV, emerges dynamically from QCD \cite{Krein:2020yor, Roberts:2021nhw, Binosi:2022djx, Papavassiliou:2022wrb, deTeramond:2022zcm, Salme:2022eoy, Ding:2022ows, Ferreira:2023fva}.  In this context of emergent hadron mass (EHM), the Nambu-Goldstone boson character of the $\pi$ and $K$ means that their properties are particularly sensitive to constructive interference between Nature's two known sources of mass, as revealed by a consideration of their mass budgets -- see, \emph{e.g}., Ref.\,\cite[Fig.\,1]{Ding:2022ows}.  This makes them important ``targets'' at foreseeable facilities
\cite{Quintans:2022utc, Brodsky:2020vco, Chen:2020ijn, Anderle:2021wcy, Arrington:2021biu, Wang:2022xad}.

Today, published data exist on $\pi$, $K$ elastic electromagnetic form factors, $F_{\pi,K}$, covering a spacelike domain that extends to $Q^2 \approx 2.5\,$GeV$^2$, where $Q^2$ is the squared momentum transfer in the process, and data being analysed reach $Q^2 \approx 8.5\,$GeV$^2$ ($\pi$) and $Q^2 \approx 5.5\,$GeV$^2$ ($K$) \cite[Table\,9.4]{Roberts:2021nhw}.  Anticipated experiments should shift these upper bounds beyond $Q^2=30\,$GeV$^2$ \cite{Brodsky:2020vco, Chen:2020ijn, Anderle:2021wcy, Arrington:2021biu}.  Moreover, whilst data relating to $\pi$ and (especially) $K$ parton distribution functions (DFs) are now scarce, that should change during the next ten years or so \cite{Adams:2018pwt, Quintans:2022utc, Brodsky:2020vco, Chen:2020ijn, Anderle:2021wcy, Arrington:2021biu, Wang:2022xad, Chang:2022pcb}.  Consequently, in the coming decade(s), one can expect to see real tests of modern predictions for these and related quantities
\cite{Gao:2017mmp, Shi:2018mcb, deTeramond:2018ecg, Chang:2020kjj, Ydrefors:2021dwa, Cui:2020tdf, Zhang:2021mtn, Raya:2021zrz, Cui:2021mom, Adhikari:2021jrh, Lu:2022cjx, dePaula:2022pcb, Albino:2022gzs, Kekez:2020vfh, Xing:2023wuk, Xing:2023pms, Lu:2023yna}.

Measurements relating to $\pi$ and $K$ generalised parton distributions (GPDs) are also likely \cite{Chen:2020ijn, Anderle:2021wcy, Arrington:2021biu, Wang:2022xad, Accardi:2023chb}.  Such data would be significant because GPDs provide access to hadron gravitational form factors \cite{Mezrag:2023nkp}; so, could open doors to comparisons between the electromagnetic and gravitational structure of $\pi$- and $K$-mesons.  This would enable insights to be drawn into the impacts of Nature's two known mass generating mechanisms on the structure of Nambu-Goldstone bosons.  It is thus imperative for theory to deliver sound, unifying predictions for $\pi$ and $K$ electromagnetic and gravitational form factors.

Steps toward a unified set of Poincar\'e invariant statements about the distributions of charge, mass, and pressure inside pions and kaons are described in Refs.\,\cite{Zhang:2021mtn, Raya:2021zrz}.  Those analyses exploit the GPD overlap representation, with light-front wave functions constrained by $\pi$ and $K$ valence-quark DFs.
Profiting from such studies, Ref.\,\cite{Xu:2023bwv} delivered a data driven prediction for the pion mass distribution, showing that the pion's mass radius is $\approx 20$\% smaller than its charge radius.

Herein, we adopt a less phenomenological approach.  Namely, using continuum Schwinger function methods (CSMs) \cite{Eichmann:2016yit, Qin:2020rad}, we solve the probe+meson scattering problem at leading-order of the truncation scheme introduced in Refs.\,\cite{Munczek:1994zz, Bender:1996bb}.  Therewith, we also unify $\pi$ and $K$ electromagnetic and gravitational distributions with predictions for an array of hadron structural properties that range over systems with up-to three heavy quarks, \emph{e.g}., Refs.\,\cite{Ding:2018xwy, Binosi:2018rht, Wang:2018kto, Qin:2019hgk, Yao:2021pdy}.

The presentation is arranged as follows.
Section~\ref{sec:CSM} introduces an approximation to the probe+meson scattering problem.
It is followed, in Sec.\,\ref{sec:vertices}, by a discussion of dressed probe+quark vertices and their properties.
Section~\ref{SecRL} details the quark+antiquark scattering kernel which underlies all concrete calculations herein.
Numerical results for the dressed probe+quark vertices are described in Sec.\,\ref{SecVertices}.
This is followed by an explanation of algebraic \emph{Ans\"atze} for all elements of the calculation, which are subsequently exploited to provide ultraviolet completions of the results.
Predictions for pion and kaon elastic electromagnetic and gravitational form factors are discussed in Sec.\,\ref{ResultsGFF}.
The gravitational form factors are used to calculate the pressure and shear force profiles drawn in Sec.\,\ref{ResultsPressure}.
Section~\ref{Epilogue} provides a summary and perspective.

\section{Form factors}
\label{sec:CSM}
\begin{figure}[t]
\centerline{\includegraphics[width=0.49\columnwidth]{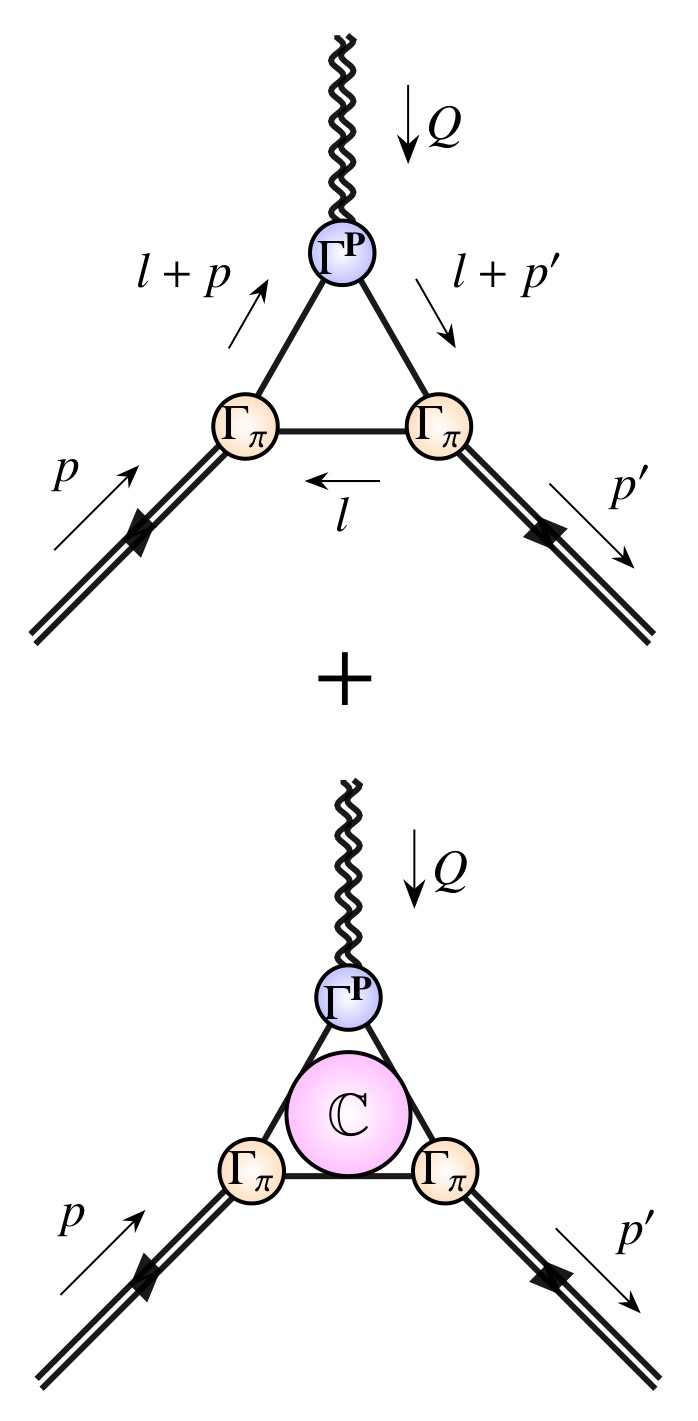}}
\caption{\label{FigCurrent}
Probe+pion interaction in RL truncation.
\emph{Both panels}.
Solid lines -- dressed quarks, $S$;
orange shaded circles -- pion Bethe-Salpeter amplitude, $\Gamma_\pi$;
and
blue shaded circles -- dressed probe+quark vertex, $\Gamma^P$.
\emph{Lower panel}.  Shaded $\mathbb C$ region -- gluon-binding contribution to the probe+pion interaction, which vanishes for $F_\pi$ but cancels a $Q$-longitudinal contribution from the top diagram when computing the graviton+pion coupling,  thereby ensuring $\bar c^\pi\equiv 0$ in Eq.\,\eqref{Lambdagpi}.
Analogous diagrams define the kaon form factors.
}
\end{figure}

Consider a charged pion, $\pi^+$, built from a $u$ valence quark and a $\bar d$ valence antiquark and suppose that isospin symmetry is exact, so the only difference between these degrees-of-freedom is their electric charge.  Then, in rain\-bow-ladder (RL) truncation \cite{Munczek:1994zz, Bender:1996bb}, the five-point Sch\-win\-ger function that defines the $\pi^+(p)\to \pi^+(p^\prime)$ elastic electromagnetic form factor takes the form drawn in the top diagram of Fig.\,\ref{FigCurrent}.  This image translates into
{\allowdisplaybreaks
\begin{subequations}
\label{pionemFF}
\begin{align}
\Lambda_\nu^{\gamma\pi}(P,Q) & = 2 N_c {\rm tr}_{D} \int \frac{d^4 l}{(2\pi)^4}
\Gamma_\nu^\gamma(l+p^\prime,l+p) L(l,P,Q)\,, \\
L(l,P,Q) & = S(l+p) \Gamma_\pi(l+p/2;p) S(l) \nonumber \\
& \quad \times \bar\Gamma_\pi(l+p^\prime/2;-p^\prime) S(l+p^\prime)\,,
\end{align}
\end{subequations}
where $2 P = p^\prime+p$, $Q=p^\prime-p$, $p^\prime\cdot p^\prime = -m_\pi^2 = p\cdot p$, with $m_\pi$ being the pion mass, $P\cdot Q=0$.  %
Further, $S(k)=1/[i \gamma\cdot k \, A(k^2)+B(k^2)]$ is the dressed-quark propagator, calculated in rainbow truncation;
$\Gamma_\pi$ is the RL-truncation pion Bethe-Salpeter amplitude;
and
$\Gamma_\nu^\gamma$ is the RL dressed photon+quark vertex.
(For an overview, see, \emph{e.g}., Refs.\,\cite{Eichmann:2016yit, Qin:2020rad}.)
Since the gluons that bind the $\pi$ do not carry electric charge, then, in RL truncation, insofar as $\Lambda_\nu^{\gamma\pi}$ is concerned, the bottom diagram in Fig.\,\ref{FigCurrent} is zero.
}

Regarding Eq.\,\eqref{pionemFF}, electromagnetic current conservation is guaranteed in RL truncation \cite{Roberts:1994hh}.  Thus, \linebreak $Q_\nu \Lambda_\nu^{\gamma\pi}(P,Q) \equiv 0$; so,
\begin{equation}
\Lambda_\nu^{\gamma\pi}(P,Q) = 2 P_\nu F_\pi(Q^2)\,,
\end{equation}
and $F_\pi(Q^2)$ can be extracted from the projection \linebreak $P_\nu \Lambda_\nu^{\gamma\pi}(P,Q)$.

The expectation value of the energy-momentum tensor in the pion, \emph{viz}.\ the $\pi$ gravitational current, takes the following form:
\begin{align}
\Lambda^g_{\mu\nu}&(P,Q) = 2 P_\mu P_\nu \theta_2^\pi(Q^2) \nonumber \\
& + \tfrac{1}{2}[Q^2 \delta_{\mu\nu} - Q_\mu Q_\nu] \theta_1^\pi(Q^2)
+ 2 m_\pi^2 \delta_{\mu\nu} \bar c^\pi(Q^2)\,,
\label{Lambdagpi}
\end{align}
where $\theta_{2,1}^\pi$ are, respectively, the in-pion mass and pressure distribution form factors, which are also accessible via the pion GPD \cite{Mezrag:2022pqk, Mezrag:2023nkp}.  The following relations follow from symmetries:
\begin{equation}
\label{symmetryresults}
\theta_2^\pi(0) = 1 \,,\quad
\theta_1^\pi(0) \stackrel{m_\pi^2=0}{=} 1 \,,
\quad \bar c^\pi(Q^2) \equiv 0\,.
\end{equation}
The first identity is a statement of mass normalisation, like $F_\pi(Q^2=0)=1$ for the electromagnetic form factor; the second is a corollary of EHM, expressed in a soft-pion theorem \cite{Polyakov:1999gs, Mezrag:2014jka}; and the third is a basic consequence of energy-momentum conservation, \emph{viz}.\ $Q_\mu \Lambda^g_{\mu\nu}(P,Q) \equiv 0 \equiv Q_\nu \Lambda^g_{\mu\nu}(P,Q)$.

In RL truncation, both diagrams in Fig.\,\ref{FigCurrent} contribute to the pion gravitational current.  The bottom diagram plays a role analogous to Ref.\,\cite[Fig.\,3B$^\prime$]{Ding:2019lwe}, which restores momentum conservation in calculations of pion DFs.  Notwithstanding that, it is not necessary herein to develop an explicit expression for this term because it only affects $\bar c^\pi(Q^2)$: $\theta_{2,1}(Q^2)$ decouple from the longitudinal projections $Q_\mu \Lambda^g_{\mu\nu}(P,Q)$, $Q_\nu \Lambda^g_{\mu\nu}(P,Q)$.  Hence, regarding $\theta_{2,1}(Q^2)$, it is sufficient to consider only the upper diagram.
This translates into an expression like Eq.\,\eqref{pionemFF}, with the only change being that the photon + quark vertex is replaced by the graviton + quark analogue: $\Gamma_\nu^\gamma(l+p^\prime,l+p) \to \Gamma_{\mu\nu}^g(l+p^\prime,l+p)$.  The gravitational form factors may then be extracted via contractions of the upper/triangle-diagram component of $\Lambda^g_{\mu\nu}(P,Q)$ with the following projectors ($L_{\mu\nu}(P) = P_{\mu}P_{\nu}/P^2$):
\begin{subequations}
\begin{align}
\mathcal{P}^{\theta_2}_{\mu \nu} & = \tfrac{1}{4 P^2} [ 3 L_{\mu\nu}(P) + L_{\mu\nu}(Q) - \delta_{\mu\nu}]\,,
\label{eq:projt2}\\
\mathcal{P}^{\theta_1}_{\mu \nu} & = \tfrac{1}{Q^2} [ - L_{\mu\nu}(P) - 3 L_{\mu\nu}(Q) + \delta_{\mu\nu}]\,.
\label{eq:projt1}
\end{align}
\end{subequations}
%

When considering $K$ form factors, the only difference from the pion is that one must distinguish quark flavours when expressing the currents illustrated by Fig.\,\ref{FigCurrent} -- see, \emph{e.g}., Refs.\,\cite{Maris:2000sk, Gao:2017mmp} for comparative studies of $\pi$ and $K$ electromagnetic form factors.  In the isospin symmetry limit, for any given form factor, $\mathscr F_K$: ${\mathscr F}_K = {\mathscr F}_K^{\bar s}+{\mathscr F}_K^u$.  (For electromagnetic form factors, one must include multiplicative quark electric charge factors.)  

We work at the hadron scale, whereat quasiparticle degrees of freedom carry all properties of a given hadron.  Evolution to higher scales \cite{Dokshitzer:1977sg, Gribov:1971zn, Lipatov:1974qm, Altarelli:1977zs}, which exposes QCD parton contributions in species-decomposi\-tions of hadron structural properties \cite{Raya:2021zrz, Lu:2022cjx}, is discussed elsewhere \cite{Cui:2021mom, Cui:2022bxn, Yin:2023dbw}.  Apart from ensuring the correct anomalous dimensions \cite{Raya:2015gva, Ding:2018xwy}, such evolution has no effect on the overall meson form factors, which are our focus.

\section{Symmetry-preserving vertices}
\label{sec:vertices}
%
%
A key factor in calculations of probe+meson form factors is the dressed probe+quark vertex.  Following thirty years of study \cite{Roberts:1994hh, Maris:1999bh, Roberts:2000aa, Xu:2021mju}, the photon+quark vertex is well understood.
%
It satisfies a Ward-Green-Takahashi (WGT) identity $(l_+^{(\prime)} = l+p^{(\prime)})$:
\begin{equation}
i Q_\nu \Gamma_\nu^\gamma(l_+^\prime,l_+)
= S^{-1}(l_+^\prime) - S^{-1}(l_+)\,.
\end{equation}
Thus, four terms in $\Gamma_\nu^\gamma$ are fixed by the dressed-quark propagator \cite{Ball:1980ay, Curtis:1990zs, Qin:2013mta}, that associated with $\sigma_{\mu\nu}l_{+\mu}^{(\prime)} l_{+\nu}$ being zero.  This is called the Ball-Chiu (BC) part of the vertex.  Eight independent nonzero terms remain.  They can be obtained by solving the inhomogeneous vector-channel Bethe-Salpeter equation using the same RL formulation as employed for $S$, $\Gamma_\pi$ in Eq.\,\eqref{pionemFF}.

Combining this information, one can decompose the vertex as follows:
\begin{align}
\Gamma_\nu^\gamma(l_+^\prime,l_+) & =
\Gamma_\nu^{\rm BC}(l_+^\prime,l_+)  \nonumber \\
& + T_{\nu\alpha}(Q)[\Gamma_\alpha^\gamma(l_+^\prime,l_+) - \Gamma_\alpha^{\rm BC}(l_+^\prime,l_+)]\,.
\label{EqPQV}
\end{align}
Here,
$T_{\nu\alpha}(Q) = \delta_{\nu\alpha} - L_{\nu\alpha}(Q)$ and,
writing $k = (l_+^\prime+l_+)/2$, $k_\pm = k \pm Q/2$ for notational convenience in Ref.\,\cite[Eq.\,(26)]{Roberts:1994hh}:
\begin{equation}
\label{BCAnsatz}
i \Gamma_\nu^{\rm BC}(k_+,k_-)
 = i \gamma_\nu \Sigma_{A_\pm} + 2 i k_\nu \gamma\cdot k \, \Delta_{A_\pm} + 2 k_\nu \Delta_{B_\pm} \,,
 %
\end{equation}
where $\Sigma_{A_\pm} = [A(k_+^2)+A(k_-^2)]/2$,
$\Delta_{F_\pm} = [F(k_+^2)-F(k_-^2)]/[k_+^2-k_-^2]$, $F=A,B$.
$\Gamma_\nu^{\rm BC}$ is a matrix-valued regular function (free of kinematic singularities for real arguments) and the second term in Eq.\,\eqref{EqPQV} exhibits, \emph{inter alia}, timelike-$Q^2$ poles, one at the mass-squared of each neutral vector meson supported by the interaction.  The lightest such state is the $\rho^0$-meson.

Equation~\eqref{EqPQV} is neither an approximation nor simplification.  Instead, it merely introduces a physical understanding into the calculation of $\Gamma_\nu^\gamma$.

%
The new element herein is the dressed graviton + quark vertex, $\Gamma_{\mu\nu}^g$, which has not hitherto been studied using a realistic gluon + quark interaction.  It is first worth recalling that $\Gamma_{\mu\nu}^g$ satisfies its own WGT identity, \emph{viz}.\ a straightforward generalisation of that for scalar fields in Ref.\,\cite{Brout:1966oea}:
\begin{equation}
\label{GvertexWGTI}
Q_\mu i\Gamma_{\mu\nu}^g(k,Q) =
S^{-1}(k_+) k_{- \nu} - S^{-1}(k_-) k_{+\nu}\,.
\end{equation}

Next, recall that $\Gamma_{\mu\nu}^g$ is an isoscalar $(I=0)$ $J^{PC}=2^{++}$ vertex.  Hence, by analogy with $\Gamma_\nu^\gamma$, $\Gamma_{\mu\nu}^g$ must exhibit timelike-$Q^2$ poles (or resonance-like structures) at the mass-squared positions of each $I=0$ meson for which a $2^{++}$ tensor coupling can be constructed.
Naturally, this includes the tower of isoscalar tensor mesons, but that does not exhaust the range of possibilities.
Poles associated with isoscalar-scalar mesons also emerge -- see, \emph{e.g}., Refs.\,\cite{Raman:1971jg, Theussl:2002xp, Xing:2022mvk}.
The lowest-mass singularity in $\Gamma_{\mu\nu}^g(k,Q)$ occurs in this channel.  It is associated with the tensor structure $Q^2 T_{\mu\nu}(Q)$ \cite{Raman:1971jg} and therefore -- see Eq.\,\eqref{Lambdagpi} -- contributes significantly to $\theta_1^\pi(Q^2)$ without much affecting $\theta_2^\pi(Q^2)$.

The trajectory of $I=0$ axialvector mesons can also contribute to $\Gamma_{\mu\nu}^g(k,Q)$.  In this case, the available tensor structures are more complex; \emph{e.g}.,  the following form is admissible:
$\epsilon_{\mu\nu\alpha\beta} \Gamma_{5\alpha}(k;Q) Q_\beta$, where $\Gamma_{5\alpha}(k;Q)$ is the inhomogeneous axialvector vertex.  However, given that $\theta_{1,2}^\pi$ are associated with $\{\mu \nu\}$-symmetric  tensors in the graviton + meson vertex and this term is antisymmetric, then axialvector mesons are not relevant in the present analysis.


Capitalising on these observations, the following minimal \emph{Ansatz} is evidently admissible as a resolution of Eq.\,\eqref{GvertexWGTI}:
\begin{align}
i\Gamma_{\mu\nu}^{g_M}&(k,Q)  \nonumber \\
& = i\Gamma_\mu^{\rm BC}(k,Q) k_\nu
-\tfrac{1}{2}\delta_{\mu\nu}[ S^{-1}(k_+) + S^{-1}(k_-)]  \nonumber \\
&
\quad +  i T_{\mu\alpha}(Q)T_{\nu\beta}(Q)4 \hat{\Gamma}^2_{\alpha\beta}(k_+,k_-)\,, \label{GvertexWGTI2}
\end{align}
where
${\Gamma}^2_{\alpha\beta}(k;Q)$ is the tensor + quark vertex generated by the inhomogeneity
\begin{equation}
\label{eq:G20}
\Gamma^{2}_{0\mu\nu}(k;Q) = T_{\mu\alpha}(Q)T_{\nu\beta}(Q) \tfrac{1}{2} \left(\gamma_\alpha k_\beta + \gamma_\beta k_\alpha \right) \,,
\end{equation}
with $\hat\Gamma(k , Q) = \Gamma(k , Q) - \Gamma(k , 0)$ so as to ensure the absence of kinematic singularities.  Notwithstanding this, dynamical singularities do appear in $\Gamma^{2}_{\mu\nu}$; namely, one at the pole position of each $I=0$ tensor meson.
Given the structure of $\mathcal{P}^{\theta_2}_{\mu \nu}$ in Eq.\,\eqref{eq:projt2}, one should expect the lowest-mass singularity in this vertex to affect $\theta_2^\pi$.  The next pole lies much deeper in the timelike region, so must play a much lesser role.
Since ${\Gamma}^2_{\mu\nu}$ possesses eight independent Dirac matrix valued tensor structures, then this \emph{Ansatz} involves fourteen such nonzero terms.

Of course, Eq.\,\eqref{GvertexWGTI2} is not complete.  Considering the available four-vectors and tensor structures, and recognising that fermion on-shell conditions cannot be used, $\Gamma_{\mu\nu}^g$ may contain a large number of additional independent terms.  Nevertheless, in general, one may write
\begin{equation}
\label{EqGQV}
\Gamma_{\mu\nu}^g(k,Q) = \Gamma_{\mu\nu}^{g_M}(k,Q) + \Gamma_{\mu\nu}^{gT}(k,Q)\,,
\end{equation}
where $\Gamma_{\mu\nu}^{gT}(k,Q)$, satisfying $Q_\mu \Gamma_{\mu\nu}^{gT}(k,Q) = 0$, represents all possible transverse structures not already included.
Plainly, like $\Gamma^{2}_{0\mu\nu}$, $\Gamma_{\mu\nu}^{gT}$ does not contribute to resolving Eq.\,\eqref{GvertexWGTI}.
It may be determined by solving the appropriate Bethe-Salpeter equation.  In doing so, one sees the emergence of isoscalar scalar mesons in the graviton+quark vertex.

Again, Eq.\,\eqref{EqGQV} is not an approximation: it just introduces a physical understanding into the calculation of $\Gamma_{\mu\nu}^g$.

\section{Interaction kernel}
\label{SecRL}
The leading-order CSM approximations to the pion electromagnetic and gravitational currents are discussed in Sec.\,\ref{sec:CSM}.  They are defined by the rainbow-ladder (RL) truncation, which is completely specified by the form of the Bethe-Salpeter kernel.  We use that introduced in Refs.\,\cite{Qin:2011dd, Binosi:2014aea}:
\begin{subequations}
\label{KDinteraction}
\begin{align}
\mathscr{K}_{\rho_1\rho_1^\prime}^{\rho_2^\prime \rho_2}(\ell) & =
\tilde{\mathpzc G}(s=\ell^2)
\tfrac{4}{3} [i\gamma_\mu]_{\rho_1\rho_1^\prime} [i\gamma_\nu]_{\rho_2^\prime\rho_2}
T_{\mu\nu}(\ell) \,,\\
%
\label{defcalG}
 \tilde{\mathpzc G}(s) & =
 \frac{8\pi^2}{\omega^4} D e^{-s/\omega^2} + \frac{8\pi^2 \gamma_m \mathcal{F}(s)}{\ln\big[ \tau+(1+s/\Lambda_{\rm QCD}^2)^2 \big]}\,,
\end{align}
\end{subequations}
where $\gamma_m=12/25$, $\Lambda_{\rm QCD} = 0.234\,$GeV, $\tau={\rm e}^2-1$, and ${\cal F}(s) = \{1 - \exp(-s/\Lambda_{\mathpzc I}^2)\}/s$, $\Lambda_{\mathpzc I}=1\,$GeV.
The origin of Eqs.\,\eqref{KDinteraction} and their links to QCD are explained elsewhere \cite{Qin:2011dd, Binosi:2014aea}.
In solving all integral equations relevant herein, we use a mass-independent (chiral-limit) momentum-subtraction renormalisation scheme \cite{Chang:2008ec}, \linebreak with renormalisation scale $\zeta=19\,$GeV$=:\zeta_{19}$.

%
%
%
%

Numerous applications have shown \cite{Ding:2022ows} that interactions in the class containing Eqs.\,\eqref{KDinteraction} can serve to unify the properties of many systems.  Contemporary studies use $\omega = 0.8\,$GeV \cite{Xu:2022kng}.  Then, with $\omega D = (0.94\,{\rm GeV})^3$ and renormalisation point invariant current quark mas\-ses (in GeV)
\begin{equation}
\label{quarkmass}
\hat m_u = \hat m_d = 0.0055\,,
\quad \hat m_s = 0.14\,,
\end{equation}
corresponding to $\hat m_s/\hat m_u =  25.5$, one obtains (in GeV):
\begin{equation}
m_\pi = 0.135\,,\; f_\pi= 0.095\,,\;
m_K = 0.495\,, \; f_K= 0.116\,,
\label{piKstatic}
\end{equation}
\emph{i.e}., values in line with experiment \cite{Workman:2022ynf}.  (Here, minor differences in comparisons with Ref.\,\cite{Xu:2022kng} owe to omission of the $[1/\ln]$-tail therein.)
Typically, if the product $\omega D$ is kept fixed, physical observables remain practically unchanged under $\omega \to (1\pm 0.2)\omega$ \cite{Qin:2020rad}.
Note that, evolved using the one-loop formula, the current masses in Eq.\,\eqref{quarkmass} correspond to $\zeta=2\,{\rm GeV}=:\zeta_2$ masses of $m_u^{\zeta_2}=3.8\,$MeV and $m_s^{\zeta_2}=97\,$MeV, respectively.

\section{Dressed vertices}
\label{SecVertices}
Having specified the Bethe-Salpeter kernel, every element in Eq.\,\eqref{pionemFF} and its analogue for the graviton+meson current, Eq.\,\eqref{Lambdagpi}, can be computed.  For instance, the photon+quark vertex is the solution of an inhomogeneous Bethe-Salpeter equation:
\begin{align}
[\Gamma_\nu^\gamma &]_{\rho_1 \rho_2}(k_+,k_-) = Z_2 [\gamma_\nu]_{\rho_1 \rho_2}
\nonumber \\
& + Z_2^2 \int_{dl}^{\Lambda} \mathscr{K}_{\rho_1\rho_1^\prime}^{\rho_2^\prime \rho_2}(k-l) [ S (l_+) \Gamma_\nu^\gamma(l_+,l_-) S (l_-)]_{\rho_1^\prime \rho_2^\prime} \,,
\label{VIBSE}
\end{align}
where
$\int_{dl}^{\Lambda}$ represents a symmetry-preserving regularisation of the four-dimensional integral, with $\Lambda$ being the regularisation scale;
$S$ is the propagator of the interacting quark, obtained from the analogous rainbow gap equation;
and $Z_2(\zeta,\Lambda)$ is the quark wave function renormalisation constant, determined as part of solving that gap equation.  ($Z_2(\zeta,\Lambda)\approx 1$ for $\Lambda /\zeta \gg 1$.)
The solution for $\Gamma_\nu^\gamma$ has eleven independent Dirac matrix valued terms and can readily be projected into the form of Eq.\,\eqref{EqPQV}.
Good numerical methods for solving sets of coupled gap and Bethe-Salpeter equations are described, \emph{e.g}., in Refs.\,\cite{Maris:1997tm, Krassnigg:2009gd}.

The remaining element in Eq.\,\eqref{Lambdagpi} is the bound-state amplitude of the meson being probed, which can be obtained from an analogous homogeneous Bethe-Salpeter equation \cite{Maris:1997tm}.


Consider now the graviton+quark vertex, the RL result for which may be obtained by solving
\begin{align}
i \Gamma_{\mu\nu}^g&(k_+,k_-)  =
Z_2 [i \gamma_\mu k_\nu
- \delta_{\mu\nu} (i\gamma\cdot k+Z_m^0 m^\zeta)] \nonumber \\
& + Z_2^2 \int_{dl}^{\Lambda} \mathscr{K}(k-l) [ S (l_+) i \Gamma_{\mu\nu}^g(l_+,l_-) S (l_-)] \,,
\label{RLGQV}
\end{align}
where $Z_m^0$ is the chiral-limit mass renormalisation constant and the WGT identity, Eq.\,\eqref{GvertexWGTI}, ensures that it is $Z_2$ which appears here.

As already noted, in general, $\Gamma_{\mu\nu}^g$ possesses a large number of independent terms.  On the other hand, we have also emphasised that, physically, one may expect just a few contributions to be important, \emph{viz}.\ those parts which saturate the WGT identity, and pieces associated with an $f_2$ tensor meson pole and an analogous scalar meson resonance.  We therefore continue with a truncated form of $\Gamma_{\mu\nu}^g$.
Specifically, the vertex given by Eq.\,\eqref{EqGQV}, with ${\Gamma}^2_{\mu\nu}$ obtained by solving the tensor analogue of Eq.\,\eqref{VIBSE} defined via the Eq.\,\eqref{eq:G20} inhomogeneity, and
\begin{equation}
 \Gamma_{\mu\nu}^{gT}(k,Q) =  T_{\mu\nu}(Q) \Gamma_{\mathbb I}(k;Q)\,,
 \label{GQVSC}
\end{equation}
where $\Gamma_{\mathbb I}(k;Q)$, which has four independent Dirac matrix valued structures, $D_{\mathbb I}^{j=1,4}\propto\{\mathbf 1, \gamma\cdot k, \gamma\cdot Q, \sigma_{\alpha\beta}k_\alpha Q_\beta\}$ -- see, \emph{e.g}., Ref.\,\cite[Appendix~A]{Krassnigg:2009zh}, is obtained by solving
\begin{align}
& {\rm tr}_D{\mathpzc P}_{\mu\nu}^j(k,Q)\Gamma_{\mu\nu}^{gT}(k_+k_-) \nonumber \\
=
& {\rm tr}_D {\mathpzc P}_{\mu\nu}^j(k,Q) Z_2^2 \int_{dl}^{\Lambda} \mathscr{K}(k-l) S (l_+) \nonumber \\
& \quad \times
 \{ \Gamma_{\mu\nu}^{g_M}(l_+,l_-) + T_{\mu\nu}(Q) \Gamma_{\mathbb I}(l_+;l_-)  \}  S (l_-) \,.
\label{GammaIRL}
\end{align}
Here,
\begin{subequations}
\begin{align}
{\rm tr}{\mathpzc P}_{\mu\nu}^j(k,Q) \Gamma_{\mu\nu}^{gT}(k,Q) & = D_{\mathbb I}^{j},\\
{\rm tr}{\mathpzc P}_{\mu\nu}^j(k,Q) \Gamma_{\mu\nu}^{g_M}(k,Q) & \equiv 0 \,.
\end{align}
\end{subequations}

Equation~\eqref{GammaIRL} is an inhomogeneous Bethe-Salpeter equation.  Its solution exhibits a pole at the mass of each scalar meson generated by the interaction in Eq.\,\eqref{KDinteraction}.  In practice and analogous to the tensor vertex, the lightest scalar dominates in the calculation of spacelike gravitational form factors because the first excitation lies roughly 1\,GeV higher in mass \cite{Xu:2022kng}, \emph{i.e}., well into the timelike region.

Solving Eq.\,\eqref{RLGQV} and the tensor analogue of Eq.\,\eqref{VIBSE}, then the lowest mass isoscalar-scalar and -tensor mesons are expressed in $\Gamma_{\mu\nu}^g$ as pole terms of the form:
\begin{equation}
\label{PolesSCAV}
\left.
\frac{-f_{\mathbb S} \Gamma^{\mathbb S}(k;Q)}{1+m_{\mathbb S}^2/Q^2}\right|_{Q^2+m_{\mathbb S}^2\simeq 0}\,,
\left. \frac{- f_{\mathbb T}
\Gamma^{\mathbb T}_{\mu\nu}(k;Q)}{1+m_{\mathbb T}^2/Q^2}\right|_{Q^2+m_{\mathbb T}^2\simeq 0}\,,
\end{equation}
where $\Gamma^{{\mathbb S},{\mathbb T}}$ are their associated canonically normalised bound-state amplitudes \cite{LlewellynSmith:1969az, Maris:1997tm}.  The calculated masses and residue coefficients are (in GeV):
\begin{equation}
\label{PoleValues}
\begin{array}{c|cccc}
    & m_{\mathbb S} & f_{\mathbb S} & m_{\mathbb T} & f_{\mathbb T} \\ \hline
%
%
u=d & 0.55 & 0.026 & 1.11 & 0.060 \\
s & 1.07 & 0.042 & 1.65 & 0.067 \\\hline
%
\end{array}\,.
\end{equation}
Here we have listed results for the probe+light-quark and probe+$s$-quark vertices.  The latter are required in the calculation of kaon form factors.

The photon+quark vertices, obtained from Eq.\,\eqref{VIBSE}, display analogous features with, \emph{e.g}.,  pole contributions from neutral vector mesons at timelike momenta \cite{Maris:1999bh, Roberts:2000aa, Xu:2021mju}.

All elements required for calculation of $\pi$ and $K$ electromagnetic and gravitational form factors are now available:
dressed quark propagators ($2$ one-variable scalar functions for each quark),
meson Bethe-Salpeter amplitudes ($4$ two-variable functions for each meson),
and probe+quark vertices ($2$ one-variable and $8$ two-variable functions for the electromagnetic interaction of each quark and $2$ one-variable and $12$ two-variable functions for the graviton+quark interaction of each quark).
The scalar functions are stored as arrays of numbers, wherewith the integrand in Eq.\,\eqref{pionemFF} -- and its analogues for other systems and probes -- can be formed and the associated integral evaluated using standard quadrature and interpolation schemes.
%

\section{Algebraic \emph{Ans\"atze}}
\label{sec:AM}
We choose to complement our fully numerical work with calculations that employ minimal effective algebraic inputs for each of the functions.  Such representations of quark propagators and meson Bethe-Salpeter amplitudes were used in Refs.\,\cite{Zhang:2021mtn, Raya:2021zrz} to deliver results for $\pi$ and $K$ GPDs and, therefrom, electromagnetic and gravitational form factors for these mesons.   Therein,
\begin{subequations}
\begin{align}
S_{q=u,s}(l) &= (-i\gamma\cdot l + M_q) /(l^2 + M_q^2) \,, \\
\label{SqlAM}
\Gamma_{{\mathscr P}=\pi,K}(l;p) & = i \gamma_5
\int_{-1}^1 dz \, \rho_{\mathscr P}(z) \, \hat\Delta(l_\omega^2,\Lambda_{\mathscr P}^2)\,,
\end{align}
\end{subequations}
where $\hat\Delta(s,u) = u /[s+u]$, $l_z = l + z p/2$, $p^2=-m_{\mathscr P}^2$,
\begin{align}
{\mathpzc n}_{\mathscr P}& \rho_{\mathscr P}(z) = \frac{1+z v_{\mathscr P}}{2 a_{\mathscr P}b_0^{\mathscr P}} \nonumber \\
& \quad \times
\left[
\mbox{sech}^2 \left(\frac{z-z_0^{\mathscr P}}{2b_0^{\mathscr P}}\right)
 +\mbox{sech}^2 \left(\frac{z+z_0^{\mathscr P}}{2b_0^{\mathscr P}}\right)\right]\,.
 \label{eq:spectralw}
 \end{align}
Our variant of the model defining parameters is listed in Table~\ref{tab:params}: the dressed-quark masses are somewhat larger than those in Refs.\,\cite[Table~I]{Raya:2021zrz} and $\Lambda_K$ is smaller.  The other values are unchanged.
(${\mathpzc n}_{{\mathscr P}=\pi,K}$ are computed normalisation constants, which ensure $F_{\mathscr P}(0)=1 = \theta_2^{\mathscr P}(0)$.)

\begin{table}[t!]
\caption{
\label{tab:params}
Used in Eqs.\,\eqref{SqlAM} -- \eqref{eq:spectralw}, one reproduces the quark propagators and meson Bethe-Salpeter amplitudes employed in Refs.\,\cite{Zhang:2021mtn, Raya:2021zrz} to computed pion and kaon GPDs.
 ($M_s = 1.35 M_u$.  Mass dimensioned quantities in GeV.)
}
\begin{center}
\begin{tabular}{l|c|c|cc|c|c|c}\hline
${\mathscr P}$ & $ m_{\mathscr P}$ & $M_u$ &
\multicolumn{2}{c|}{$M_{\bar h} \equiv \Lambda_{\mathscr P}$}  & $b_0^{\mathscr P}$ & $z_0^{\mathscr P}$ & $v_{\mathscr P}$  \\
\hline
$\pi_{h=d}$ & $0.135$ & 0.38 & $M_u$ & $M_u$ & $0.316$ & $1.23\phantom{5}$ & 0\phantom{.41} \\
$K_{h=s}$ &  $0.495$ & 0.38 & $M_s$   & $M_s$  & $0.1\phantom{75}$ & $0.625$ & $0.41$ \\\hline
\end{tabular}
\end{center}
\end{table}

Using Eq.\,\eqref{SqlAM}, then Eq.\,\eqref{BCAnsatz} simplifies to $\Gamma_\nu^{\rm BC} = \gamma_\nu$; hence, we use $\Gamma_\nu^\gamma = \gamma_\nu$ as the algebraic representation.

Somewhat more care must be taken with the graviton + quark vertex owing to the importance of scalar- and tensor-meson poles.  Considering Eqs.\,\eqref{GvertexWGTI2}\,--\,\eqref{EqGQV}, \linebreak \eqref{GQVSC}, \eqref{PolesSCAV}, 
we are led to the following \emph{Ans\"atze}
\begin{align}
\Gamma_{\mu\nu}^{gq}&(k;Q)  = i \delta_{\mu\nu} [ i \gamma\cdot k + M_q]
+ \gamma_\mu k_\nu \nonumber \\
%
& + T_{\mu\alpha}(Q) \gamma_\alpha k_\nu 4 P^{\mathbb T}_q(Q^2)
+  T_{\mu\nu}(Q) {\mathbf 1}  P^{{\mathbb S}}_q(Q^2)\,,
\end{align}
where, on $Q^2\in [0,5]\,$GeV$^2$,  
\begin{subequations}
\label{modelBSAs}
\begin{align}
P_q^{\mathbb T}(t) & = 
 \frac{- t}{t+m_{{\mathbb T}_q}^2 }
\frac{m_{{\mathbb T}_q}^2 (1-\kappa_q)^2}{t + \kappa_q^2 m_{{\mathbb T}_q}^2}\,, \\
%
P_q^{{\mathbb S}}(t) & = \frac{ - t{\mathpzc r}_{{\mathbb S}_q} }{t+m_{{\mathbb S}_q}^2} \,.
\end{align}
\end{subequations}
%
Extension to $Q^2 \gtrsim 5\,$GeV$^2$ is discussed in Sec.\,\ref{ResultsPressure}.
Fixing the masses in Eq.\,\eqref{modelBSAs} to be those computed in RL truncation, listed in Eq.\,\eqref{PoleValues}, then each graviton+quark vertex \emph{Ansatz} has two parameters:
$\kappa_q$,
${\mathpzc r}_{{\mathbb S}_q}$.

Regarding $\kappa_q$, we desire to implement a mass scale, $\kappa_q m_{{\mathbb T}_q}$,  in the form factor that is associated with the tensor channel.  Having already used the ground state mass for the leading pole term, we chose $\kappa_u=\kappa_s =: \kappa = \,$ the ratio of masses of the first radial excitation and ground state in the light-quark channel.  Using the listings in Ref.\,\cite[RPP]{Workman:2022ynf}, one finds a value of $1.13$ for this ratio; so, we proceed with $\kappa = 1.13$, a fixed value, \emph{viz}.\ it is not subsequently varied.  Using this approach, one finds that the residues of the tensor poles in $P^{\mathbb T}_{u,s}$ are (in GeV):
$f_{{\mathbb T}_u} = 0.046$,
$f_{{\mathbb T}_s} = 0.059$.
Given the simplicity of the \emph{Ans\"atze}, these values are a fair match with the results in Eq.\,\eqref{PoleValues}.

Having fixed $\kappa$, the sole variable parameter in each algebraic graviton+quark vertex \emph{Ansatz} is ${\mathpzc r}_{{\mathbb S}_q}$.  As noted above, scalar meson poles contribute significantly to $\theta_1^{\mathscr P}$ without materially affecting $\theta_2^{\mathscr P}$.  Further, any deviation from the chiral-limit result, \emph{viz}.\ $\theta_1^{\mathscr P}(0)=1$, see Eq.\,\eqref{symmetryresults}, derives from quark current-mass corrections \cite{Polyakov:2018zvc, Mezrag:2014jka, Xing:2022mvk}.  In fact, one can use chiral effective field theory to estimate $\theta_1^{\pi}(0)=0.97(1)$ \cite{Polyakov:2018zvc}.
%
Using our algebraic \emph{Ans\"atze}, this central value is obtained with ${\mathpzc r}_{{\mathbb S}_u}=0.089\,$GeV, which translates into $f_{{\mathbb S}_u}=0.012\,$GeV.

The value of ${\mathpzc r}_{{\mathbb S}_s}$ can be fixed once $\theta_1^{K}(0)$ is known.  Owing to the character of interference between EHM and HB-generated current masses, then, in stepping away from the chiral limit, SU$(3)$ flavour symmetry breaking is normally expressed in observables by an amount of the order $f_\pi/f_K$ or $f_\pi^2/f_K^2$ \cite{Cui:2020tdf, Zhang:2021mtn, Raya:2021zrz, Roberts:2021nhw}.   So, one may anticipate $\theta_1^{K}(0) \approx 0.77(10)$.
Harmonising with this, chiral effective field theory yields $\theta_1^{K}(0) \approx 0.77(15)$ \cite{Polyakov:2018zvc}.
The central value, $\theta_1^{K}(0) = 0.77$, is obtained with ${\mathpzc r}_{{\mathbb S}_s} \approx 0.32\,$GeV, which corresponds to $f_{{\mathbb S}_s}=0.039\,$GeV.

Again, given the simplicity of the \emph{Ans\"atze}, the $f_{{\mathbb S}_{u,s}}$ values compare tolerably with the RL results in Eq.\,\eqref{PoleValues}.

A common way to estimate uncertainties in RL predictions is to reevaluate all results with $\pm 5$\% variations of $\omega$ in Eq.\,\eqref{KDinteraction}.  Given the correlation between $\omega$ and derived mass scales, we translate this approach into an estimation of uncertainties via simultaneous $\pm 5$\% variations of each mass scale in the algebraic \emph{Ans\"atze}, \emph{viz}.\ $M_q$, $m_{{\mathbb S}_q}$, $m_{{\mathbb T}_q}$.  The resulting change in a given value is the uncertainty listed in each instance below.


\section{Electromagnetic and gravitational form factors}
\label{ResultsGFF}
Using the RL results, described in Secs.\,\ref{SecRL}, \ref{SecVertices}, for all elements in the photon+meson current, Eq.\,\eqref{pionemFF}, and its analogue for the graviton+meson current, one can directly proceed to deliver a unified set of predictions for $\pi$ and $K$ electromagnetic and gravitational form factors.  The first studies of $F_{\pi,K}(Q^2)$ were completed in Ref.\,\cite{Maris:2000sk}.  Using brute-force numerical techniques, as therein, then in the calculation of each form factor one encounters moving singularities in the complex-$l^2$ domain sampled by the bound-state equations \cite{Maris:1997tm} such that there is a maximum value of $Q^2$ beyond which the evaluation of integrals like that in Eqs.\,\eqref{pionemFF} is no longer possible with conventional algorithms.

More advanced methods have been developed \cite{Raya:2015gva, Gao:2017mmp, Ding:2018xwy}, exploiting the perturbation theory integral representation (PTIR) \cite{Nakanishi:1969ph}.  However, constructing accurate PTIRs is time consuming.  This is especially true in our case because one would need to build PTIRs for each quark propagator, Bethe-Salpeter amplitude, photon+quark and graviton+quark vertex considered herein, \emph{i.e}., roughly 200 scalar functions.
We therefore continue with a straightforward RL approach, computing all form factors on the directly accessible domain and then using the algebraic \emph{Ans\"atze}, detailed in Sec.\,\ref{sec:AM}, to assist in defining their ultraviolet completions.

\begin{figure}[t]
\vspace*{0ex}

\leftline{\hspace*{0.5em}{{\textsf{A}}}}
\vspace*{-2ex}
\centerline{\includegraphics[width=0.85\columnwidth]{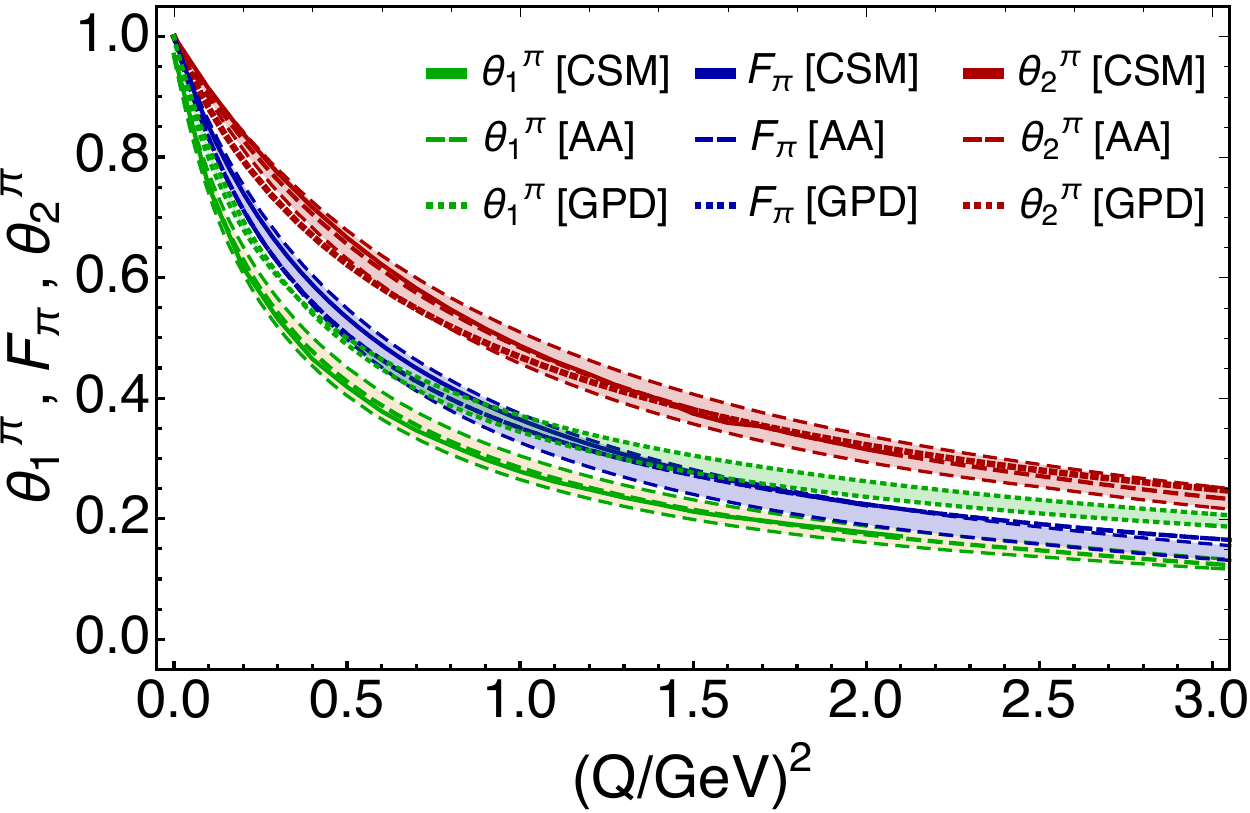}}
\vspace*{2ex}

\leftline{\hspace*{0.5em}{{\textsf{B}}}}
\vspace*{-2ex}
\centerline{\includegraphics[width=0.85\columnwidth]{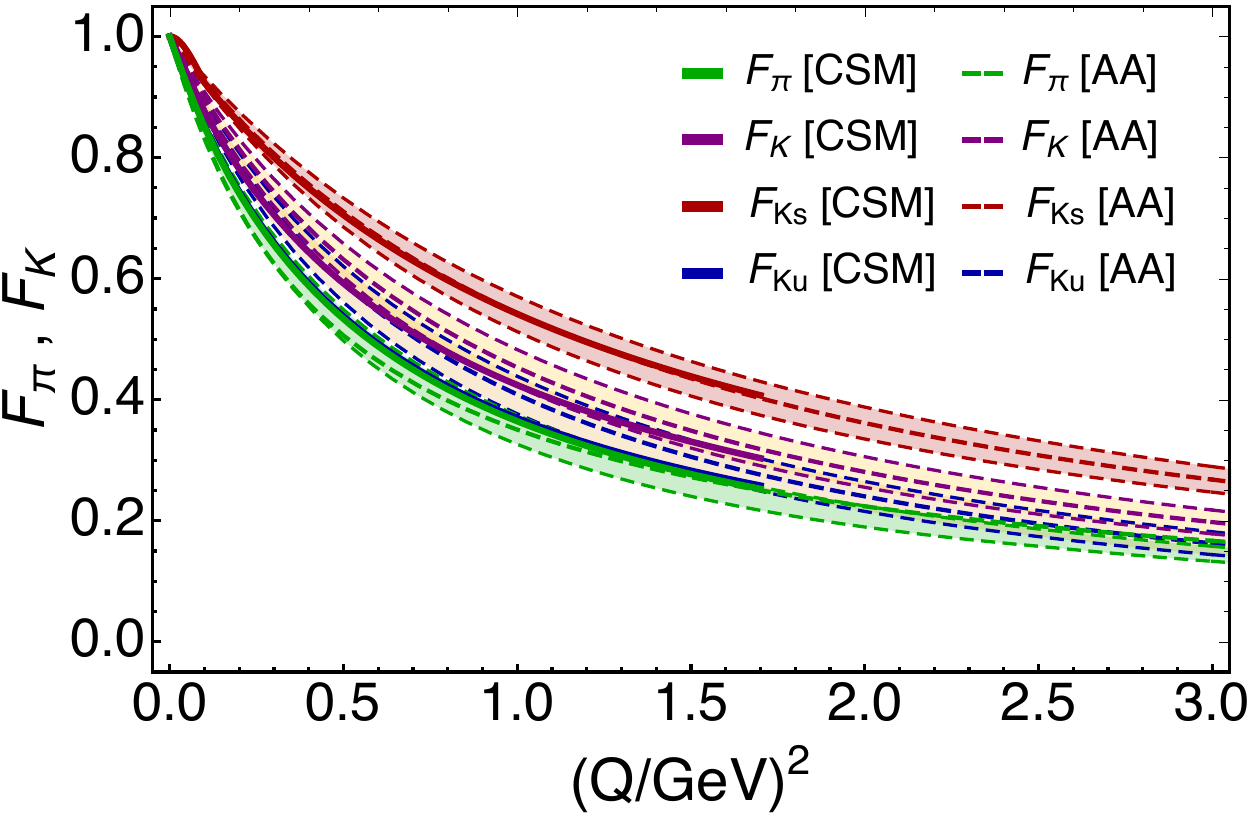}}

\caption{
\emph{Panel A}.
Pion electromagnetic ($F_\pi$ - blue)
and gravitational form factors ($\theta_1$ - green, $\theta_2$ - red).
Legend. CSM predictions, Secs.\,\ref{SecRL}, \ref{SecVertices} -- solid curves;
algebraic \emph{Ans\"atze}, Sec.\,\ref{sec:AM} -- dashed curves;
pion GPD \emph{Ans\"atze} \cite{Raya:2021zrz} -- dotted curves.
\label{fig:GFFsPi}
\emph{Panel B}.
Kaon electromagnetic form factors.
Legend.
As in Panel A, with
$K$ total -- purple; $\bar s$ in $K$ -- red; and $u$ in $K$ -- blue.
$F_\pi$ comparison curves -- green.
}
\end{figure}

\begin{table}[t!]
\caption{
\label{tab:results}
CSM (RL truncation - Secs.\,\ref{SecRL}, \ref{SecVertices}) predictions for various pseudoscalar meson static properties and comparisons with other selected calculations (AA - Sec.\,\ref{sec:AM}; GPD \emph{Ans\"atze} \cite{Raya:2021zrz}).
As usual, the radius associated with a given form factor,
$\mathscr F(t)$, $\mathscr F(0)\neq 0$, is obtained via $r^2 = -\left. 6 (d/dt) \ln \mathscr F(t) \right|_{t=0}$.
Recent analyses of data yield $r_\pi^{\theta_2} = 0.51(2)$ \cite{Xu:2023bwv}, $r_\pi^F=0.64(2)$ \cite{Cui:2021aee} and $r_\pi^{\theta_2}/r_\pi^F=0.79(3)$ \cite{Xu:2023bwv}, $r_K^F \approx 0.53$ \cite{Cui:2021aee}.
(All radii, $r_{\mathscr P}$, listed in fm.)
}
\begin{center}
\begin{tabular*}
{\hsize}
{
l@{\extracolsep{0ptplus1fil}}
|l@{\extracolsep{0ptplus1fil}}
l@{\extracolsep{0ptplus1fil}}
l@{\extracolsep{0ptplus1fil}}
l@{\extracolsep{0ptplus1fil}}}\hline
        & $\theta_1^{\mathscr P}(0)$ & $r_{\mathscr P}^{\theta_1}$ & $r_{\mathscr P}^{F}$ & $r_{\mathscr P}^{\theta_2}$ \\\hline
$\pi_{\rm CSM}$ & $0.97$  & $0.81\phantom{(3)}$ & $0.64\phantom{(3)}$ & $0.47\phantom{(3)}$ \\
$\pi_{\rm AA}$ & $0.97$  & $0.80(4)$ & $0.64(3)$ & $0.49(3)$ \\
$\pi_{\rm GPD}$ &  & $0.81\phantom{(3)}$ & $0.69\phantom{(3)}$ & $ 0.56\phantom{(3)}$ \\\hline
$K_{\rm CSM}$ & $0.77$  & $0.63\phantom{(3)}$ & $0.58\phantom{(3)}$ & $0.40\phantom{(3)}$ \\
$K_{\rm AA}$ & $0.77$  & $0.68(4)$ & $0.51(3)$ & $0.41(3)$ \\
$K_{\rm CSM}^{\bar s}$ & $0.43$  & $0.39\phantom{(3)}$ & $0.44\phantom{(3)}$ & $0.37\phantom{(3)}$ \\
$K_{\rm AA}^{\bar s}$ & $0.43$  & $0.42(3)$ & $0.43(3)$ & $0.36(3)$ \\
$K_{\rm CSM}^u$ & $0.34$  & $0.85\phantom{(3)}$ & $0.64\phantom{(3)}$ & $0.45\phantom{(3)}$ \\
$K_{\rm AA}^u$ & $0.34$  & $0.91(5)$ & $0.55(4)$ & $0.48(3)$ \\
\hline
\end{tabular*}
\end{center}
\end{table}

Our CSM (RL truncation) predictions for pion elastic electromagnetic and gravitational form factors are drawn in Fig.\,\ref{fig:GFFsPi}A.
In a direct calculation, such as ours, which avoids the GPD route to gravitational form factors, there is no $D$-term ambiguity in the analysis of $\theta_1(t)$ \cite{Chouika:2017rzs}.
It is notable, therefore, that the value $\theta_1(t=0)=0.97$ -- see Table~\ref{fig:GFFsPi}, matches the estimate obtained using chiral effective field theory \cite{Polyakov:2018zvc}.
The predictions for the $\theta_2^\pi$ and $F_\pi$ radii are consistent with recent extractions \cite{Xu:2023bwv, Cui:2021aee}, confirming thereby that the distribution of mass within the pion is more compact than the distribution of charge: $r_\pi^{\theta_2}/r_\pi^{F} = 0.74$.  Moreover, the pressure distribution radius is greater than both the electromagnetic and mass radii: $r_\pi^{\theta_1} > r_\pi^{F} > r_\pi^{\theta_2}$, with $r_\pi^{F}/r_\pi^{\theta_1} \approx 0.79$.
Our predictions for these gravitational radii are in accord with values inferred from measurements of $\gamma^\ast \gamma \to \pi^0\pi^0$ \cite{Kumano:2017lhr}.

As evident in Fig.\,\ref{fig:GFFsPi}A, using simple algorithms, the $\pi$ RL calculation fails on $Q^2 \gtrsim 2\,$GeV$^2$.  Thereupon, the algebraic \emph{Ans\"atze}, Sec.\,\ref{sec:AM}, become valuable.  On $Q^2\lesssim 2\,$GeV$^2$, they deliver results in agreement with the RL truncation.  This justifies their use in developing ultraviolet completions of the RL predictions.

The results in Ref.\,\cite{Raya:2021zrz} were obtained using algebraic \emph{Ans\"atze} for pion and kaon generalised parton distributions (GPDs), constrained entirely by hadron-scale $\pi$ and $K$ valence-parton DFs.  Owing to the $D$-term ambiguity \cite{Chouika:2017rzs} encountered, \emph{e.g}., in connecting GPDs with gravitational form factors, reliable results for $\theta_1^{\mathscr P}(0)$ could not be obtained in Ref.\,\cite{Raya:2021zrz}.  Herein, there is no $D$-term ambiguity: as already seen for the pion, the RL truncation supplies a definite prediction for $\theta_1^{\mathscr P}(t)$.

\begin{figure}[t]
\vspace*{0ex}

\leftline{\hspace*{0.5em}{{\textsf{A}}}}
\vspace*{-2ex}
\centerline{\includegraphics[width=0.85\columnwidth]{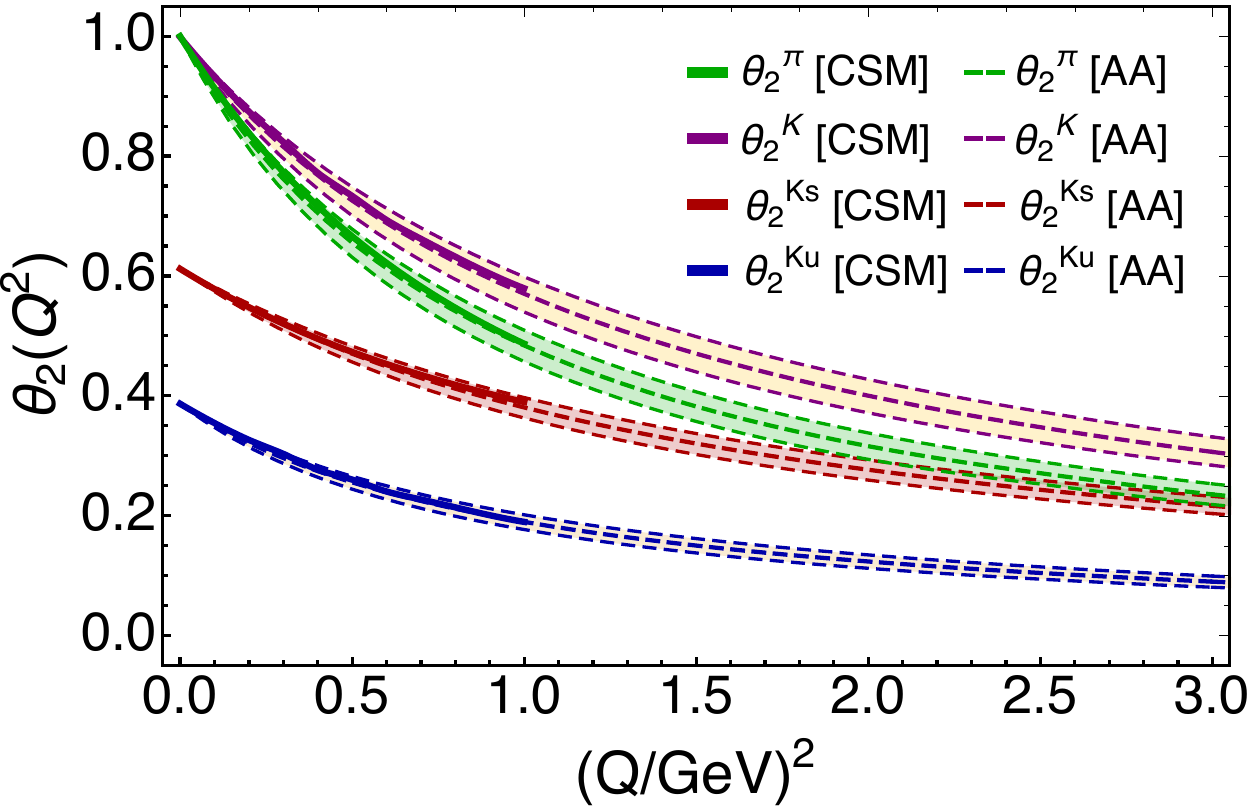}}
\vspace*{2ex}

\leftline{\hspace*{0.5em}{{\textsf{B}}}}
\vspace*{-2ex}
\centerline{\includegraphics[width=0.85\columnwidth]{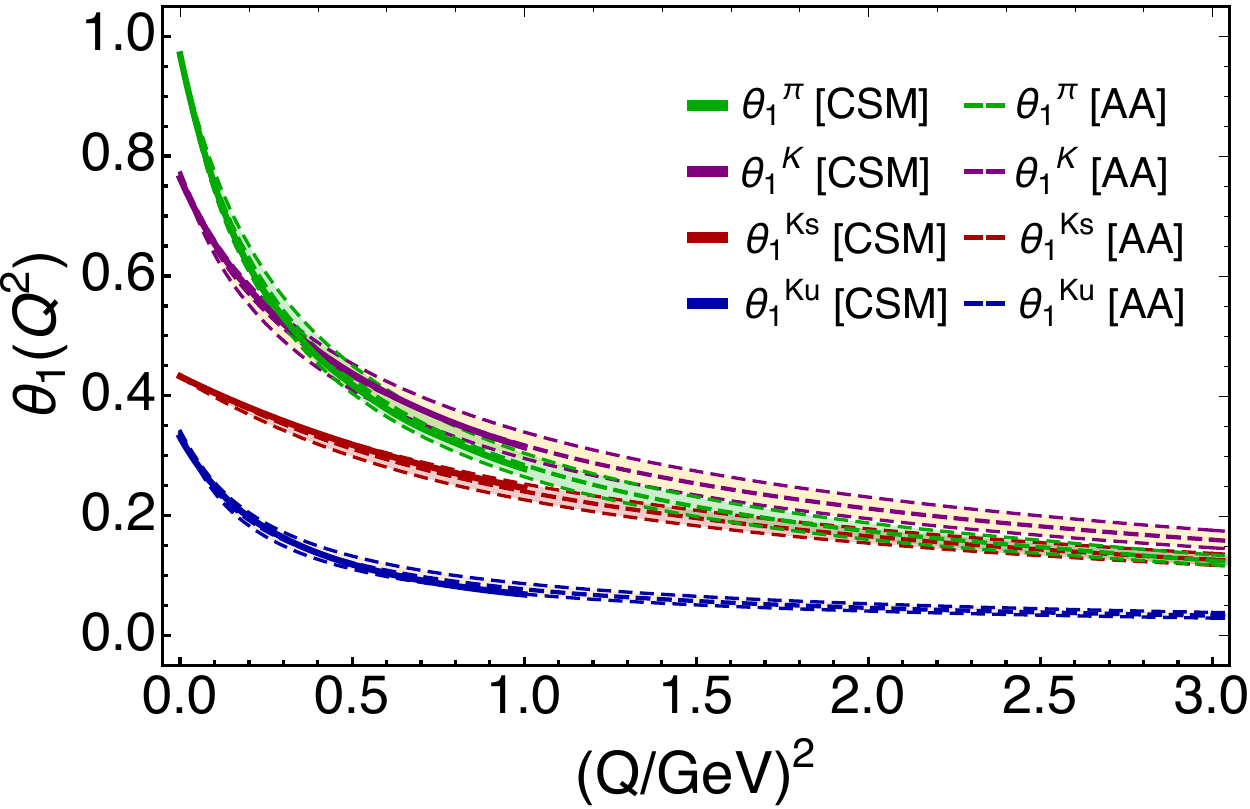}}

\caption{\label{fig:GFFsKB}
Kaon gravitational form factors.
\emph{Panel A}.  Mass, $\theta_2^K$.
\emph{Panel B}.  Pressure, $\theta_1^K$.
Legend.
CSM predictions - solid curves; and results from algebraic \emph{Ans\"atze} - dashed curves.
Kaon total -- purple; $\bar s$ quark in $K$ -- red; and $u$ quark in $K$ -- blue.
Pion comparison curves -- green.
}
\end{figure}

The CSM predictions for kaon elastic electromagnetic and gravitational form factors are drawn in Figs.\,\ref{fig:GFFsPi}B, \ref{fig:GFFsKB}.  These figures show that, for the $K$, the RL calculation fails on $Q^2 \gtrsim 1.7\,$GeV$^2$.  The algebraic \emph{Ans\"atze}, Sec.\,\ref{sec:AM}, are valuable on this domain: since they deliver results in agreement with RL truncation on $Q^2\lesssim 1.7\,$GeV$^2$, they can be used to assist in defining ultraviolet completions of the RL predictions.

Figure~\ref{fig:GFFsPi}B shows that the $u$ quark in $K$ elastic electric form factor is almost indistinguishable from the $u$ quark in $\pi$ form factor, $F_\pi$.  This weak environmental sensitivity is typical of RL results for light pseudoscalar meson elastic electromagnetic form factors \cite{Maris:2000sk}.  On the other hand, as seen elsewhere \cite{Gao:2017mmp}, there is a marked difference between the $\bar s$ in $K$ and $u $ in $K$ electromagnetic form factors: $F_K^{\bar s}(Q^2)$, $F_K^{u}(Q^2)$, respectively.  A similar distinction is expressed in the ratio of $\bar s$ in $K$ and $u $ in $K$ valence parton DFs \cite{Badier:1980jq, Cui:2020tdf, Lin:2020ssv}.

Table~\ref{fig:GFFsPi} reveals that RL truncation yields $\theta_1^K(0)=0.77$, a value which matches the estimate discussed in closing Sec.\,\ref{sec:AM}, thereby confirming the character of EHM + HB interference in elastic form factors.  In addition, $\theta_1^{K_u}(0)/\theta_1^{K_{\bar s}}(0)=0.79$, which may be compared with our calculated value of $f_\pi/f_K = 0.82$, Eq.\,\eqref{piKstatic}, and the empirical result $f_\pi/f_K = 0.84$ \cite{Workman:2022ynf}.

On the other hand, regarding Fig.\,\ref{fig:GFFsKB}A, RL truncation is seen to predict $\theta_2^{K_u}(0) = 0.39$, $\theta_2^{K_s}(0) = 0.61$.  These values are consistent with gravitational current conservation, Eq.\,\eqref{symmetryresults}; however, they overestimate the effects of flavour splitting.
Indeed, one should have $\theta_2^{K_q}(0)  = \langle x \rangle_{\zeta_{\cal H}}^{q_K}$, \emph{i.e}., the light-front momentum fraction carried by the $q$ valence quark in the kaon at the hadron scale, $\zeta_{\cal H}$; and modern analyses of kaon DFs yield \cite{Cui:2020tdf}:
$\langle x \rangle_{\zeta_{\cal H}}^{u_K}=0.47(1)$,
$\langle x \rangle_{\zeta_{\cal H}}^{{\bar s}_K}=0.53(1)$.
These are the results of CSM calculations that incorporate effects beyond RL truncation.
Working with information available in Ref.\,\cite{Shi:2014uwa} and exploiting connections elucidated in Ref.\,\cite[Sec.\,3A]{Roberts:2021nhw}, analogous RL predictions are
$\langle x \rangle_{\zeta_{\cal H}}^{u_K}=0.42(1)$,
$\langle x \rangle_{\zeta_{\cal H}}^{{\bar s}_K}=0.58(1)$.

These comparisons confirm the character of the results for $\theta_2^{K_{u,{\bar s}}}(0)$ in Fig.\,\ref{fig:GFFsKB}A; namely, the accentuated magnitude of flavour splitting is an artefact of RL truncation.
The issue can be traced to the fact that an efficacious RL truncation must express all effects of EHM in the form of $\tilde {\mathpzc G}(s)$ -- see Eq.\,\eqref{KDinteraction}, achieving a description of hadron phenomena by overmagnifying the gauge-sector interaction strength on $s\lesssim \Lambda_{\mathpzc I}^2$  \cite{Binosi:2014aea}.  Moving beyond RL truncation, EHM is also manifested in corrections to the gluon+quark vertex \cite{Binosi:2016wcx}, thereby enabling an explanation of observables with a far less infrared enhanced form of $\tilde {\mathpzc G}$; hence, more realistic expressions of EHM+HB interference.

Further regarding the gravitational form factors, the ordering of kaon radii is the same as that found for the pion:
$r_K^{\theta_1} > r_K^{F} > r_K^{\theta_2}$ -- see Table~\ref{tab:results}.
Moreover, in each case, the net kaon radius is smaller than the kindred pion radius.  In fact, averaging over all form factors: $r_K / r_\pi = 0.85(6)$.
Here, matching the electromagnetic form factor pattern, the $\theta_K^{1,2}$ $u$ quark in $K$ radii are very like the analogous $u$ in $\pi$ radii: $r_{K_u}^{\theta_1} \approx 1.09 r_{\pi}^{\theta_1}$; $r_{\pi}^{\theta_2}  \approx 1.03 r_{K_u}^{\theta_2}$.

\section{Pressure profiles}
\label{ResultsPressure}
Working with the Poincar\'e-invariant elastic form factors described above, one may compute Breit frame charge, mass, pressure, and shear force distributions via appropriate three dimensional Fourier transforms \cite{Polyakov:2018zvc}.  Other transforms are possible \cite{Freese:2022fat}; but since the input function in any case is always the same, then whatever type of transform is chosen, it is just a mathematical operation on the same input object.  Thus, judiciously interpreted, all outputs are qualitatively equivalent.

Calculation of a Fourier transform requires knowledge of the subject form factor on $Q^2\in [0,\infty)$.  However, our direct CSM predictions do not extend beyond $Q^2 = 2\,$GeV$^2$.  The algebraic \emph{Ans\"atze} discussed in Sec.\,\ref{sec:AM} can be used to assist in developing their ultraviolet completions.  Those results are readily obtained on $Q^2\in [0,5]\,$GeV$^2$; but in order to arrive at realistic pictures, they must be augmented by inclusion of additional $\ln Q^2$ suppression at ultraviolet momenta.  This is necessary to express the fact that, in four spacetime dimensions, a quantum field theory treatment of hadron form factors introduces such scaling violations.  Neglecting the extra suppression, the transforms yield spurious short-distance divergences -- see, \emph{e.g}., Ref.\,\cite[Fig.\,4]{Zhang:2020ecj}.

The required completions can be achieved by fitting the form factor results obtained using the algebraic \emph{Ans\"atze} with the following function:
\begin{equation}\label{eq:largeQ2}
{\mathscr F}(y=Q^2/\Lambda_{\mathpzc I}^2)= {\mathscr F}_0 \frac{1+b_1 y}{1+b_2 y + b_3 y^2} \frac{1+a_0 y}{1+a_0 y \ln{\left(1+a_0 y\right)}} \,.
\end{equation}
where ${\mathscr F}_0 := {\mathscr F}(y=0)$.  The coefficients are listed in Table~\ref{FitParams}.

\begin{table}[t]
    \caption{Inserted into Eq.\,\eqref{eq:largeQ2}, these parameters define realistic ultraviolet completions of the $\pi$, $K$ electromagnetic and gravitational form factors.
    \label{FitParams}}
\begin{center}
\begin{tabular*}
{\hsize}
{
l@{\extracolsep{0ptplus1fil}}
|l@{\extracolsep{0ptplus1fil}}
l@{\extracolsep{0ptplus1fil}}
l@{\extracolsep{0ptplus1fil}}
l@{\extracolsep{0ptplus1fil}}
l@{\extracolsep{0ptplus1fil}}}\hline
            & ${\mathscr F}_0$ & $b_1$ & $b_2$ & $b_3$ & $a_0$ \\
        \hline
        $F_\pi$ & 1 & 0.055 & 1.886 & 0.259 & 0.046 \\[0.4ex]
        $F_K^u$ & 1 & 0.066 & 1.492 & 0.394 & 0.121 \\[0.4ex]
        $F_K^{\bar{s}}$ & 1 & 0.162 & 0.991 & 0.253 & 0.050 \\
        \hline
       $\theta_2^\pi$ & 1 & 0.316 & 1.376 & 0.434 & 0.035 \\[0.4ex]
        $\theta_2^{K_u}$ & 0.387 & 0.318 & 1.479 & 0.558 & 0.155 \\[0.4ex]
        $\theta_2^{K_{\bar{s}}}$ & 0.613 & 1.293 & 1.990 & 1.114 & 0.131 \\
    \hline
           $\theta_1^\pi$ & 0.969 & 19.10 & 22.35 & 55.20 & 0.184 \\[0.4ex]
        $\theta_1^{K_u}$ & 0.331 & 15.04 & 18.68 & 57.54 & 0.137 \\[0.4ex]
        $\theta_1^{K_{\bar{s}}}$ & 0.436 & \;\;1.536 & \;\;2.318 & \;\;1.719 & 0.118 \\
        \hline
    \end{tabular*}
    \end{center}
\end{table}

Charge and mass distribution Breit frame density profiles, calculated from ${\mathscr F} = F_{\mathscr P}, \theta_2^{\mathscr P}$, respectively, are fairly straightforward.  In all cases,
$r^2 \rho_{\mathscr F}(r)$ vanishes at $r=0$;
and with increasing $r$,
grows to a maximum value (at $r \sim 0.2 - 0.4$),
whose location is closer to $r=0$ and magnitude is greater for form factors with smaller radii;
then approaches zero from above at a rate that increases as the form factor radius becomes smaller.
Naturally, consistent with the results in Table~\ref{tab:results}, the density profile of a meson's heavier valence quark is more compact than that of the lighter valence quark.

\begin{figure}[t]
\vspace*{0ex}

\leftline{\hspace*{0.5em}{{\textsf{A}}}}
\vspace*{-2ex}
\centerline{\includegraphics[width=0.85\columnwidth]{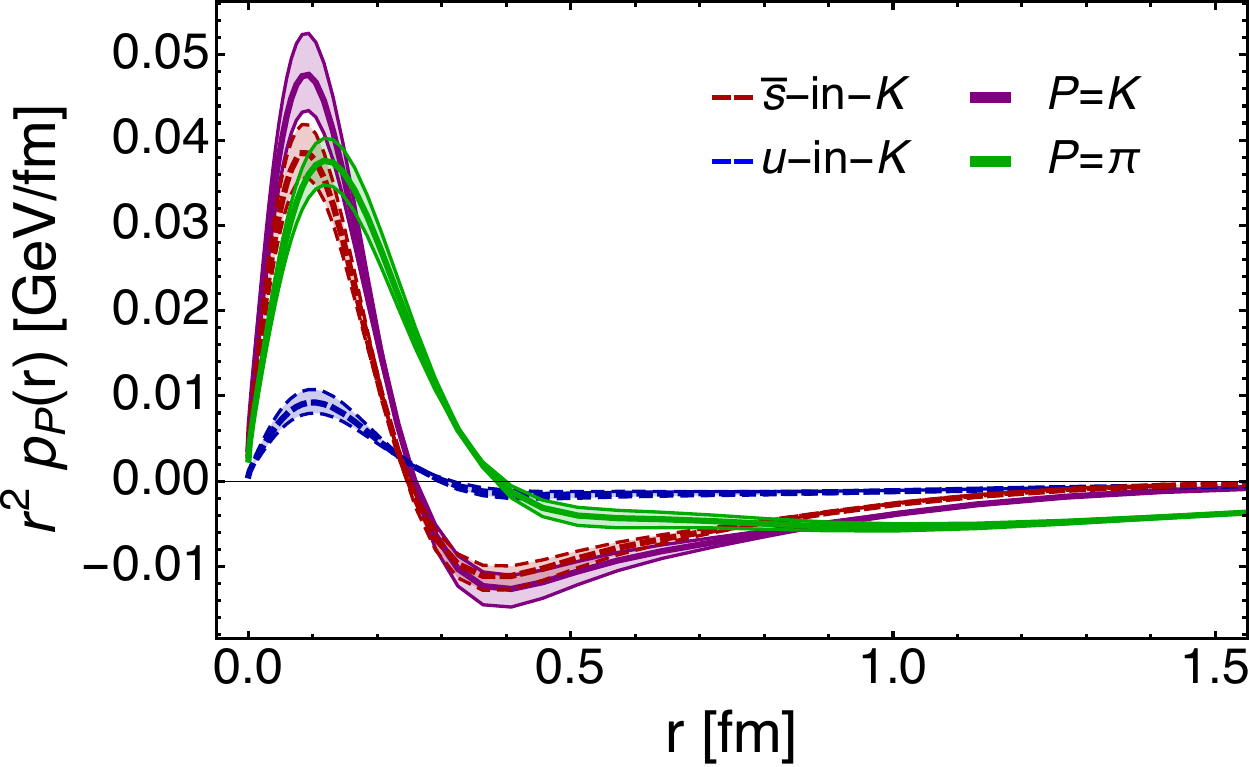}}
\vspace*{2ex}

\leftline{\hspace*{0.5em}{{\textsf{B}}}}
\vspace*{-2ex}
\centerline{\includegraphics[width=0.85\columnwidth]{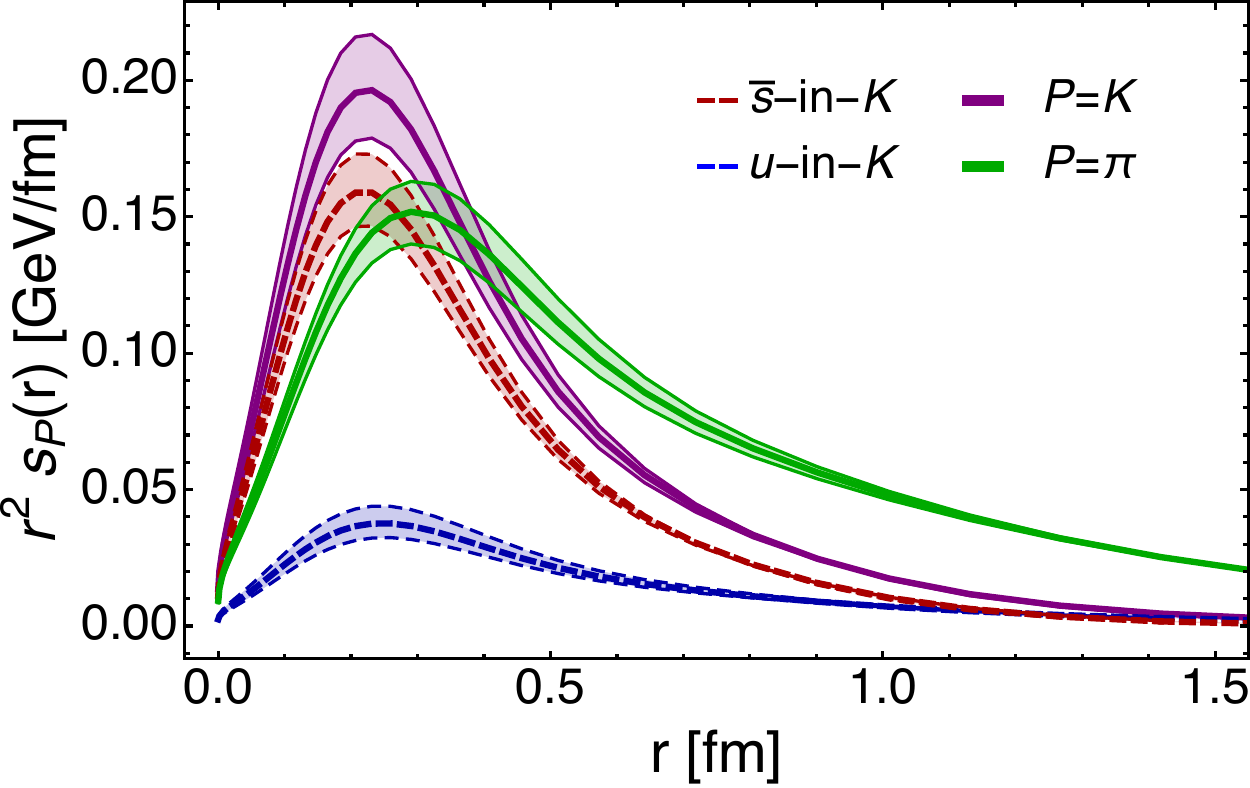}}

\caption{\label{FigPressures}
Pressure (\emph{Panel A}) and shear (\emph{Panel B}) distributions obtained using Eqs.\,\eqref{EqPressure} as explained in Sec.\,\ref{ResultsPressure}.
Legend.
$\pi$ -- solid green curves;
$K$ -- solid purple curves;
$\bar s$ in $K$ -- dashed red curves;
and
$u$ in $K$ -- dashed blue curves.
}
\end{figure}

Pseudoscalar meson pressure and shear force distributions may be defined as follows \cite{Polyakov:2018zvc}:
{\allowdisplaybreaks
\begin{subequations}
\label{EqPressure}
\begin{align}
p_{\mathscr P}^q(r)  & =
 \frac{1}{6\pi^2 r} \int_0^\infty d\Delta \,\frac{\Delta}{2 E(\Delta)} \, \sin(\Delta r) [\Delta^2\theta_1^{{\mathscr P}_q}(\Delta^2)] \,, \label{EqPressureA}\\
 s_{\mathscr P}^q (r)  & =
%
\frac{3}{8 \pi^2} \int_0^\infty d\Delta \,\frac{\Delta^2}{2 E(\Delta)} \, {\mathpzc j}_2(\Delta r) \, [\Delta^2\theta_1^{{\mathscr P}_q}(\Delta^2)] \,, \label{EqPressureB}
\end{align}
\end{subequations}
where
isospin symmetry is assumed,
$q=u,s$,
$\Delta = \surd Q^2$,
$E(\Delta)^2=m_{\mathscr P}^2+\Delta^2/4$
and ${\mathpzc j}_2(z)$ is a spherical Bessel function,
For bound states, $\int_0^\infty dr\,r^2 p_{\mathscr P}^q(r)  \equiv 0$.
As revealed by Fig.\,\ref{FigPressures}, these densities possess more interesting structure than the charge and mass profiles.  Semiquantitatively, our ultraviolet completed RL predictions yield the same pictures obtained using the GPD models detailed in Refs.\,\cite{Zhang:2021mtn, Raya:2021zrz}.
}

Consider first the pressure profiles, drawn in Fig.\,\ref{FigPressures}A.
Compared with the GPD model results \cite{Zhang:2021mtn, Raya:2021zrz}, these profiles are more compact because the RL predictions for $\theta_1^{{\mathscr P}_q}$ fall faster with increasing $\Delta$.
The qualitative features of the profiles in Fig.\,\ref{FigPressures}A suggest a physical interpretation.  Namely, the meson pressures are positive and large on the neighbourhood $r\simeq 0$, whereupon the meson's dressed-valence constituents are pushing away from each other.  With increasing separation, the pressure switches sign, indicating a transition to the domain wherewithin confinement forces exert their influence on the pair.  The zeros occur at the following locations (in fm):
$r_c^\pi=0.39(1)$, $r_c^K=0.26(1)$, $r_c^{K_{u}}=0.30(1)$, $r_c^{K_{\bar s}}=0.25(1)$.
%


As noted elsewhere \cite{Zhang:2020ecj, Zhang:2021mtn, Raya:2021zrz}, profiles like those in Fig.\,\ref{FigPressures}A can be drawn for neutron stars.  They indicate $r\simeq  0$ pressures therein of roughly $0.1\,$GeV/fm \cite{Ozel:2016oaf}.  Plainly, 
the form factors computed herein yield core pressures in Nambu-Goldstone bosons that are similar in magnitude to that of a neutron star.

The $K$ profile is more compact, so the associated peak core pressure is higher than in the $\pi$: the ratio is $\approx 1.5$.  Further, 
the $\bar s$-quark contributes more of the total $K$ pressure than its partner $u$-quark and its peak/trough intensities are greater.

The shear pressures, drawn in Fig.\,\ref{FigPressures}B, are an indicator of the strength of deformation forces within the meson.  Evidently, these forces are maximal in the neighbourhood upon which the pressure changes sign.  Within these neighbourhoods, the forces driving the quark and antiquark apart are overwhelmed by attractive confinement pressure.  It is interesting to compare the total $\pi$ and $K$ shear forces:
\begin{equation}
\label{Totals}
\int_0^\infty dr\,r^2 s_{K}(r) = 0.77 \int_0^\infty dr\,r^2 s_{\pi}(r)\,.
\end{equation}
The $\pi$ result is greater owing to the long tail of $s_{\pi}(r)$: on $r\in[0,0.9]$, the integrated strengths are equal.  The ordering in Eq.\,\eqref{Totals} is opposite to that found using the GPD models in Refs.\,\cite{Zhang:2021mtn, Raya:2021zrz}.  This difference owes largely to (\emph{a}) the failure of those models to express the suppression of $\theta_1^K(0)$ -- see Table~\ref{tab:results}; and (\emph{b}) a low value for $r_{K_u}^{\theta_1}$.

%

\section{Summary and perspective}
\label{Epilogue}
Using continuum Schwinger function methods, we delivered predictions for $\pi$ and $K$ elastic electromagnetic, $F_{\pi,K}$, and gravitational, $\theta_{1,2}^{\pi,K}$, form factors, thereby unifying them with numerous additional properties of these Nambu-Goldstone bosons and other hadrons.  The calculations were completed using a sym\-metry-preser\-ving
formulation of each quantum field equation 
relevant to the calculation of the bound-state wave functions and interaction currents.  In particular, we described and calculated the structure of dressed graviton + quark vertices [Secs.\,\ref{sec:vertices}, \ref{SecVertices}].  The analysis highlights that, just as the low momentum transfer behaviour of hadron electromagnetic form factors is sensitive to the properties of the lightest neutral vector meson in a given quark+antiquark channel, so their mass and pressure form factors are sensitive, respectively, to the analogous tensor  and scalar mesons.

%
For both $\pi$ and $K$, the pressure distribution, $\theta_1$, is softer than $F$, which is softer than the mass distribution, $\theta_2$; and each $K$ form factor is harder than its kindred $\pi$ form factor [Sec.\,\ref{ResultsGFF}].
Examining form factor flavour separations, the sizes of the effects are typical of those found when considering interference between Nature's two known sources of mass, \emph{i.e}., emergent hadron mass and Higgs-boson couplings into QCD.

Regarding density and pressure profiles -- Sec.\,\ref{ResultsPressure}, defined as form factor Fourier transforms, each follows the expected pattern, \emph{e.g}., $K$ profiles are more compact than those of the $\pi$, and the $K$ core pressure is higher.
Our direct calculations, which do not suffer from the $D$-term ambiguity encountered in GPD-based analyses, confirm that the pressure at the core of a Nambu-Goldstone boson is of the same magnitude as that in the heart of a neutron star.
Considering the pressure profiles, one might define a confinement radius as the location at which these functions cross zero.  We find this radius to be $\approx 0.39\,$fm for the pion.  The kaon value is roughly 33\% smaller, with the $\bar s$ in $K$ radius being $\approx 15$\% smaller than that of the partner $u$ quark.
Contrary to earlier phenomenological results, the total shear pressure in the $\pi$ exceeds that in the $K$.

The analysis described herein should enable a direct calculation of nucleon gravitational form factors using either a quark+diquark picture \cite{Barabanov:2020jvn} or a three-body approach \cite{Eichmann:2011vu, Wang:2018kto}.

\medskip
\noindent\textbf{Acknowledgments}.
We are grateful for constructive remarks from L.~Chang and Z.-F.~Cui.
Work supported by:
National Natural Science Foundation of China (grant nos.\,12135007, 12233002);
%
Helmholtz-Zentrum Dresden Rossendorf, under the High Potential Programme;
Spanish Ministry of Science and Innovation (MICINN) (grant no.\ PID2022-140440NB-C22);
Junta de Andaluc{\'{\i}}a (grant no.\ P18-FR-5057);
and
STRONG-2020 ``The strong interaction at the frontier of knowledge: fundamental research and applications” which received funding from the European Union's Horizon 2020 research and innovation programme (grant agreement no.\ 824093).


\medskip

\noindent\textbf{Data Availability Statement} This manuscript has no associated data
or the data will not be deposited. [Authors' comment: All information necessary to reproduce the results described herein is contained in the material presented above.]


\begin{thebibliography}{92}
\providecommand{\natexlab}[1]{#1}
\providecommand{\url}[1]{\texttt{#1}}
\providecommand{\urlprefix}{URL }
\expandafter\ifx\csname urlstyle\endcsname\relax
  \providecommand{\doi}[1]{doi:\discretionary{}{}{}#1}\else
  \providecommand{\doi}[1]{doi:\discretionary{}{}{}\begingroup
  \urlstyle{rm}\url{#1}\endgroup}\fi
\providecommand{\bibinfo}[2]{#2}

\bibitem[{Adams et~al.(2018)}]{Adams:2018pwt}
\bibinfo{author}{B.~Adams}, et~al., \bibinfo{title}{{Letter of Intent: A New
  QCD facility at the M2 beam line of the CERN SPS (COMPASS++/AMBER) --
  arXiv:1808.00848 [hep-ex]$\!$}} .

\bibitem[{Brodsky et~al.(2020)}]{Brodsky:2020vco}
\bibinfo{author}{S.~J. Brodsky}, et~al., \bibinfo{title}{{Strong QCD from
  Hadron Structure Experiments}}, \bibinfo{journal}{Int. J. Mod. Phys. E}
  \bibinfo{volume}{29}~(\bibinfo{number}{08}) (\bibinfo{year}{2020})
  \bibinfo{pages}{2030006}.

\bibitem[{Chen et~al.(2020)Chen, Guo, Roberts, and Wang}]{Chen:2020ijn}
\bibinfo{author}{X.~Chen}, \bibinfo{author}{F.-K. Guo}, \bibinfo{author}{C.~D.
  Roberts}, \bibinfo{author}{R.~Wang}, \bibinfo{title}{{Selected Science
  Opportunities for the EicC}}, \bibinfo{journal}{Few Body Syst.}
  \bibinfo{volume}{61} (\bibinfo{year}{2020}) \bibinfo{pages}{43}.

\bibitem[{Anderle et~al.(2021)}]{Anderle:2021wcy}
\bibinfo{author}{D.~P. Anderle}, et~al., \bibinfo{title}{{Electron-ion collider
  in China}}, \bibinfo{journal}{Front. Phys. (Beijing)}
  \bibinfo{volume}{16}~(\bibinfo{number}{6}) (\bibinfo{year}{2021})
  \bibinfo{pages}{64701}.

\bibitem[{Arrington et~al.(2021)}]{Arrington:2021biu}
\bibinfo{author}{J.~Arrington}, et~al., \bibinfo{title}{{Revealing the
  structure of light pseudoscalar mesons at the electron\textendash{}ion
  collider}}, \bibinfo{journal}{J. Phys. G} \bibinfo{volume}{48}
  (\bibinfo{year}{2021}) \bibinfo{pages}{075106}.

\bibitem[{Quintans(2022)}]{Quintans:2022utc}
\bibinfo{author}{C.~Quintans}, \bibinfo{title}{{The New AMBER Experiment at the
  CERN SPS}}, \bibinfo{journal}{Few Body Syst.}
  \bibinfo{volume}{63}~(\bibinfo{number}{4}) (\bibinfo{year}{2022})
  \bibinfo{pages}{72}.

\bibitem[{Wang and Chen(2022)}]{Wang:2022xad}
\bibinfo{author}{R.~Wang}, \bibinfo{author}{X.~Chen}, \bibinfo{title}{{The
  Current Status of Electron Ion Collider in China}}, \bibinfo{journal}{Few
  Body Syst.} \bibinfo{volume}{63}~(\bibinfo{number}{2}) (\bibinfo{year}{2022})
  \bibinfo{pages}{48}.

\bibitem[{Chang et~al.(2023)Chang, Peng, Platchkov, and Sawada}]{Chang:2022pcb}
\bibinfo{author}{W.-C. Chang}, \bibinfo{author}{J.-C. Peng},
  \bibinfo{author}{S.~Platchkov}, \bibinfo{author}{T.~Sawada},
  \bibinfo{title}{{Fixed-target charmonium production and pion parton
  distributions}}, \bibinfo{journal}{Phys. Rev. D}
  \bibinfo{volume}{107}~(\bibinfo{number}{5}) (\bibinfo{year}{2023})
  \bibinfo{pages}{056008}.

\bibitem[{Carman et~al.(2023)Carman, Gothe, Mokeev, and
  Roberts}]{Carman:2023zke}
\bibinfo{author}{D.~S. Carman}, \bibinfo{author}{R.~W. Gothe},
  \bibinfo{author}{V.~I. Mokeev}, \bibinfo{author}{C.~D. Roberts},
  \bibinfo{title}{{Nucleon Resonance Electroexcitation Amplitudes and Emergent
  Hadron Mass}}, \bibinfo{journal}{Particles}
  \bibinfo{volume}{6}~(\bibinfo{number}{1}) (\bibinfo{year}{2023})
  \bibinfo{pages}{416--439}.

\bibitem[{Accardi et~al.(2023)}]{Accardi:2023chb}
\bibinfo{author}{A.~Accardi}, et~al., \bibinfo{title}{{Strong Interaction
  Physics at the Luminosity Frontier with 22 GeV Electrons at Jefferson Lab --
  arXiv:2306.09360 [nucl-ex]}} .

\bibitem[{Krein and Peixoto(2020)}]{Krein:2020yor}
\bibinfo{author}{G.~Krein}, \bibinfo{author}{T.~C. Peixoto},
  \bibinfo{title}{{Femtoscopy of the Origin of the Nucleon Mass}},
  \bibinfo{journal}{Few Body Syst.} \bibinfo{volume}{61}~(\bibinfo{number}{4})
  (\bibinfo{year}{2020}) \bibinfo{pages}{49}.

\bibitem[{Roberts et~al.(2021)Roberts, Richards, Horn, and
  Chang}]{Roberts:2021nhw}
\bibinfo{author}{C.~D. Roberts}, \bibinfo{author}{D.~G. Richards},
  \bibinfo{author}{T.~Horn}, \bibinfo{author}{L.~Chang},
  \bibinfo{title}{{Insights into the emergence of mass from studies of pion and
  kaon structure}}, \bibinfo{journal}{Prog. Part. Nucl. Phys.}
  \bibinfo{volume}{120} (\bibinfo{year}{2021}) \bibinfo{pages}{103883}.

\bibitem[{Binosi(2022)}]{Binosi:2022djx}
\bibinfo{author}{D.~Binosi}, \bibinfo{title}{{Emergent Hadron Mass in Strong
  Dynamics}}, \bibinfo{journal}{Few Body Syst.}
  \bibinfo{volume}{63}~(\bibinfo{number}{2}) (\bibinfo{year}{2022})
  \bibinfo{pages}{42}.

\bibitem[{Papavassiliou(2022)}]{Papavassiliou:2022wrb}
\bibinfo{author}{J.~Papavassiliou}, \bibinfo{title}{{Emergence of mass in the
  gauge sector of QCD}}, \bibinfo{journal}{Chin. Phys. C}
  \bibinfo{volume}{46}~(\bibinfo{number}{11}) (\bibinfo{year}{2022})
  \bibinfo{pages}{112001}.

\bibitem[{de~Teramond(2022)}]{deTeramond:2022zcm}
\bibinfo{author}{G.~F. de~Teramond}, \bibinfo{title}{{Emergent phenomena in
  QCD: The holographic perspective -- arXiv:2212.14028 [hep-ph]}}, in:
  \bibinfo{booktitle}{{25th Workshop on What Comes Beyond the Standard
  Models?}}, \bibinfo{year}{2022}.

\bibitem[{Salm\`e(2022)}]{Salme:2022eoy}
\bibinfo{author}{G.~Salm\`e}, \bibinfo{title}{{Explaining mass and spin in the
  visible matter: the next challenge}}, \bibinfo{journal}{J. Phys. Conf. Ser.}
  \bibinfo{volume}{2340}~(\bibinfo{number}{1}) (\bibinfo{year}{2022})
  \bibinfo{pages}{012011}.

\bibitem[{Ding et~al.(2023)Ding, Roberts, and Schmidt}]{Ding:2022ows}
\bibinfo{author}{M.~Ding}, \bibinfo{author}{C.~D. Roberts},
  \bibinfo{author}{S.~M. Schmidt}, \bibinfo{title}{{Emergence of Hadron Mass
  and Structure}}, \bibinfo{journal}{Particles}
  \bibinfo{volume}{6}~(\bibinfo{number}{1}) (\bibinfo{year}{2023})
  \bibinfo{pages}{57--120}.

\bibitem[{Ferreira and Papavassiliou(2023)}]{Ferreira:2023fva}
\bibinfo{author}{M.~N. Ferreira}, \bibinfo{author}{J.~Papavassiliou},
  \bibinfo{title}{{Gauge Sector Dynamics in QCD}}, \bibinfo{journal}{Particles}
  \bibinfo{volume}{6}~(\bibinfo{number}{1}) (\bibinfo{year}{2023})
  \bibinfo{pages}{312--363}.

\bibitem[{Gao et~al.(2017)Gao, Chang, Liu, Roberts, and Tandy}]{Gao:2017mmp}
\bibinfo{author}{F.~Gao}, \bibinfo{author}{L.~Chang}, \bibinfo{author}{Y.-X.
  Liu}, \bibinfo{author}{C.~D. Roberts}, \bibinfo{author}{P.~C. Tandy},
  \bibinfo{title}{{Exposing strangeness: projections for kaon electromagnetic
  form factors}}, \bibinfo{journal}{Phys. Rev. D}
  \bibinfo{volume}{96}~(\bibinfo{number}{3}) (\bibinfo{year}{2017})
  \bibinfo{pages}{034024}.

\bibitem[{Shi et~al.(2018)Shi, Mezrag, and Zong}]{Shi:2018mcb}
\bibinfo{author}{C.~Shi}, \bibinfo{author}{C.~Mezrag}, \bibinfo{author}{H.-S.
  Zong}, \bibinfo{title}{{Pion and kaon valence quark distribution functions
  from Dyson-Schwinger equations}}, \bibinfo{journal}{Phys. Rev. D}
  \bibinfo{volume}{98} (\bibinfo{year}{2018}) \bibinfo{pages}{054029}.

\bibitem[{de~Teramond et~al.(2018)de~Teramond, Liu, Sufian, Dosch, Brodsky, and
  Deur}]{deTeramond:2018ecg}
\bibinfo{author}{G.~F. de~Teramond}, \bibinfo{author}{T.~Liu},
  \bibinfo{author}{R.~S. Sufian}, \bibinfo{author}{H.~G. Dosch},
  \bibinfo{author}{S.~J. Brodsky}, \bibinfo{author}{A.~Deur},
  \bibinfo{title}{{Universality of Generalized Parton Distributions in
  Light-Front Holographic QCD}}, \bibinfo{journal}{Phys. Rev. Lett.}
  \bibinfo{volume}{120} (\bibinfo{year}{2018}) \bibinfo{pages}{182001}.

\bibitem[{Chang et~al.(2020)Chang, Raya, and Wang}]{Chang:2020kjj}
\bibinfo{author}{L.~Chang}, \bibinfo{author}{K.~Raya},
  \bibinfo{author}{X.~Wang}, \bibinfo{title}{{Pion Parton Distribution Function
  in Light-Front Holographic QCD}}, \bibinfo{journal}{Chin. Phys. C}
  \bibinfo{volume}{44}~(\bibinfo{number}{11}) (\bibinfo{year}{2020})
  \bibinfo{pages}{114105}.

\bibitem[{Ydrefors et~al.(2021)Ydrefors, de~Paula, Nogueira, Frederico, and
  Salm\'e}]{Ydrefors:2021dwa}
\bibinfo{author}{E.~Ydrefors}, \bibinfo{author}{W.~de~Paula},
  \bibinfo{author}{J.~H.~A. Nogueira}, \bibinfo{author}{T.~Frederico},
  \bibinfo{author}{G.~Salm\'e}, \bibinfo{title}{{Pion electromagnetic form
  factor with Minkowskian dynamics}}, \bibinfo{journal}{Phys. Lett. B}
  \bibinfo{volume}{820} (\bibinfo{year}{2021}) \bibinfo{pages}{136494}.

\bibitem[{Cui et~al.(2020)Cui, Ding, Gao, Raya, Binosi, Chang, Roberts,
  Rodr\'{\i}guez-Quintero, and Schmidt}]{Cui:2020tdf}
\bibinfo{author}{Z.-F. Cui}, \bibinfo{author}{M.~Ding},
  \bibinfo{author}{F.~Gao}, \bibinfo{author}{K.~Raya},
  \bibinfo{author}{D.~Binosi}, \bibinfo{author}{L.~Chang},
  \bibinfo{author}{C.~D. Roberts},
  \bibinfo{author}{J.~Rodr\'{\i}guez-Quintero}, \bibinfo{author}{S.~M.
  Schmidt}, \bibinfo{title}{{Kaon and pion parton distributions}},
  \bibinfo{journal}{Eur. Phys. J. C} \bibinfo{volume}{80}
  (\bibinfo{year}{2020}) \bibinfo{pages}{1064}.

\bibitem[{Zhang et~al.(2021{\natexlab{a}})Zhang, Raya, Chang, Cui, Morgado,
  Roberts, and Rodr\'\i{}guez-Quintero}]{Zhang:2021mtn}
\bibinfo{author}{J.-L. Zhang}, \bibinfo{author}{K.~Raya},
  \bibinfo{author}{L.~Chang}, \bibinfo{author}{Z.-F. Cui},
  \bibinfo{author}{J.~M. Morgado}, \bibinfo{author}{C.~D. Roberts},
  \bibinfo{author}{J.~Rodr\'\i{}guez-Quintero}, \bibinfo{title}{{Measures of
  pion and kaon structure from generalised parton distributions}},
  \bibinfo{journal}{Phys. Lett. B} \bibinfo{volume}{815}
  (\bibinfo{year}{2021}{\natexlab{a}}) \bibinfo{pages}{136158}.

\bibitem[{Raya et~al.(2022)Raya, Cui, Chang, Morgado, Roberts, and
  Rodr{\'{\i}}guez-Quintero}]{Raya:2021zrz}
\bibinfo{author}{K.~Raya}, \bibinfo{author}{Z.-F. Cui},
  \bibinfo{author}{L.~Chang}, \bibinfo{author}{J.-M. Morgado},
  \bibinfo{author}{C.~D. Roberts},
  \bibinfo{author}{J.~Rodr{\'{\i}}guez-Quintero}, \bibinfo{title}{{Revealing
  pion and kaon structure via generalised parton distributions}},
  \bibinfo{journal}{Chin. Phys. C} \bibinfo{volume}{46}~(\bibinfo{number}{26})
  (\bibinfo{year}{2022}) \bibinfo{pages}{013105}.

\bibitem[{Cui et~al.(2022{\natexlab{a}})Cui, Ding, Morgado, Raya, Binosi,
  Chang, Papavassiliou, Roberts, Rodr\'\i{}guez-Quintero, and
  Schmidt}]{Cui:2021mom}
\bibinfo{author}{Z.~F. Cui}, \bibinfo{author}{M.~Ding}, \bibinfo{author}{J.~M.
  Morgado}, \bibinfo{author}{K.~Raya}, \bibinfo{author}{D.~Binosi},
  \bibinfo{author}{L.~Chang}, \bibinfo{author}{J.~Papavassiliou},
  \bibinfo{author}{C.~D. Roberts},
  \bibinfo{author}{J.~Rodr\'\i{}guez-Quintero}, \bibinfo{author}{S.~M.
  Schmidt}, \bibinfo{title}{{Concerning pion parton distributions}},
  \bibinfo{journal}{Eur. Phys. J. A} \bibinfo{volume}{58}~(\bibinfo{number}{1})
  (\bibinfo{year}{2022}{\natexlab{a}}) \bibinfo{pages}{10}.

\bibitem[{Adhikari et~al.(2021)Adhikari, Mondal, Nair, Xu, Jia, Zhao, and
  Vary}]{Adhikari:2021jrh}
\bibinfo{author}{L.~Adhikari}, \bibinfo{author}{C.~Mondal},
  \bibinfo{author}{S.~Nair}, \bibinfo{author}{S.~Xu}, \bibinfo{author}{S.~Jia},
  \bibinfo{author}{X.~Zhao}, \bibinfo{author}{J.~P. Vary},
  \bibinfo{title}{{Generalized parton distributions and spin structures of
  light mesons from a light-front Hamiltonian approach}},
  \bibinfo{journal}{Phys. Rev. D} \bibinfo{volume}{104}~(\bibinfo{number}{11})
  (\bibinfo{year}{2021}) \bibinfo{pages}{114019}.

\bibitem[{Lu et~al.(2022)Lu, Chang, Raya, Roberts, and
  Rodr\'\i{}guez-Quintero}]{Lu:2022cjx}
\bibinfo{author}{Y.~Lu}, \bibinfo{author}{L.~Chang}, \bibinfo{author}{K.~Raya},
  \bibinfo{author}{C.~D. Roberts},
  \bibinfo{author}{J.~Rodr\'\i{}guez-Quintero}, \bibinfo{title}{{Proton and
  pion distribution functions in counterpoint}}, \bibinfo{journal}{Phys. Lett.
  B} \bibinfo{volume}{830} (\bibinfo{year}{2022}) \bibinfo{pages}{137130}.

\bibitem[{de~Paula et~al.(2022)de~Paula, Ydrefors, Nogueira~Alvarenga,
  Frederico, and Salm\`e}]{dePaula:2022pcb}
\bibinfo{author}{W.~de~Paula}, \bibinfo{author}{E.~Ydrefors},
  \bibinfo{author}{J.~H. Nogueira~Alvarenga}, \bibinfo{author}{T.~Frederico},
  \bibinfo{author}{G.~Salm\`e}, \bibinfo{title}{{Parton distribution function
  in a pion with Minkowskian dynamics}}, \bibinfo{journal}{Phys. Rev. D}
  \bibinfo{volume}{105}~(\bibinfo{number}{7}) (\bibinfo{year}{2022})
  \bibinfo{pages}{L071505}.

\bibitem[{Albino et~al.(2022)Albino, Higuera-Angulo, Raya, and
  Bashir}]{Albino:2022gzs}
\bibinfo{author}{L.~Albino}, \bibinfo{author}{I.~M. Higuera-Angulo},
  \bibinfo{author}{K.~Raya}, \bibinfo{author}{A.~Bashir},
  \bibinfo{title}{{Pseudoscalar mesons: Light front wave functions, GPDs, and
  PDFs}}, \bibinfo{journal}{Phys. Rev. D}
  \bibinfo{volume}{106}~(\bibinfo{number}{3}) (\bibinfo{year}{2022})
  \bibinfo{pages}{034003}.

\bibitem[{Kekez and Klabu\v{c}ar(2023)}]{Kekez:2020vfh}
\bibinfo{author}{D.~Kekez}, \bibinfo{author}{D.~Klabu\v{c}ar},
  \bibinfo{title}{{Pion observables calculated in Minkowski and Euclidean
  spaces with Ans\"atze for quark propagators}}, \bibinfo{journal}{Phys. Rev.
  D} \bibinfo{volume}{107}~(\bibinfo{number}{9}) (\bibinfo{year}{2023})
  \bibinfo{pages}{094025}.

\bibitem[{Xing et~al.(2023{\natexlab{a}})Xing, Ding, Cui, Pimikov, Roberts, and
  Schmidt}]{Xing:2023wuk}
\bibinfo{author}{H.~Y. Xing}, \bibinfo{author}{M.~Ding}, \bibinfo{author}{Z.~F.
  Cui}, \bibinfo{author}{A.~V. Pimikov}, \bibinfo{author}{C.~D. Roberts},
  \bibinfo{author}{S.~M. Schmidt}, \bibinfo{title}{{Constraining the pion
  distribution amplitude using Drell-Yan reactions on a proton --
  arXiv:2308.13695 [hep-ph]}} .

\bibitem[{Xing et~al.(2023{\natexlab{b}})Xing, Yao, Li, Binosi, Cui, and
  Roberts}]{Xing:2023pms}
\bibinfo{author}{H.-Y. Xing}, \bibinfo{author}{Z.-Q. Yao},
  \bibinfo{author}{B.-L. Li}, \bibinfo{author}{D.~Binosi},
  \bibinfo{author}{Z.-F. Cui}, \bibinfo{author}{C.~D. Roberts},
  \bibinfo{title}{{Developing predictions for pion fragmentation functions --
  arXiv:2311.01613 [hep-ph]}} .

\bibitem[{Lu et~al.(2023)Lu, Xu, Raya, Roberts, and
  Rodr\'\i{}guez-Quintero}]{Lu:2023yna}
\bibinfo{author}{Y.~Lu}, \bibinfo{author}{Y.-Z. Xu}, \bibinfo{author}{K.~Raya},
  \bibinfo{author}{C.~D. Roberts},
  \bibinfo{author}{J.~Rodr\'\i{}guez-Quintero}, \bibinfo{title}{{Pion
  distribution functions from low-order Mellin moments -- arXiv:2311.08565
  [hep-ph]}} .

\bibitem[{Mezrag(2023)}]{Mezrag:2023nkp}
\bibinfo{author}{C.~Mezrag}, \bibinfo{title}{{Generalised Parton Distributions
  in Continuum Schwinger Methods: Progresses, Opportunities and Challenges}},
  \bibinfo{journal}{Particles} \bibinfo{volume}{6}~(\bibinfo{number}{1})
  (\bibinfo{year}{2023}) \bibinfo{pages}{262--296}.

\bibitem[{Xu et~al.(2023{\natexlab{a}})Xu, Raya, Cui, Roberts, and
  Rodr\'\i{}guez-Quintero}]{Xu:2023bwv}
\bibinfo{author}{Y.-Z. Xu}, \bibinfo{author}{K.~Raya}, \bibinfo{author}{Z.-F.
  Cui}, \bibinfo{author}{C.~D. Roberts},
  \bibinfo{author}{J.~Rodr\'\i{}guez-Quintero}, \bibinfo{title}{{Empirical
  Determination of the Pion Mass Distribution}}, \bibinfo{journal}{Chin. Phys.
  Lett. \emph{Express}} \bibinfo{volume}{40}~(\bibinfo{number}{4})
  (\bibinfo{year}{2023}{\natexlab{a}}) \bibinfo{pages}{041201}.

\bibitem[{Eichmann et~al.(2016)Eichmann, Sanchis-Alepuz, Williams, Alkofer, and
  Fischer}]{Eichmann:2016yit}
\bibinfo{author}{G.~Eichmann}, \bibinfo{author}{H.~Sanchis-Alepuz},
  \bibinfo{author}{R.~Williams}, \bibinfo{author}{R.~Alkofer},
  \bibinfo{author}{C.~S. Fischer}, \bibinfo{title}{{Baryons as relativistic
  three-quark bound states}}, \bibinfo{journal}{Prog. Part. Nucl. Phys.}
  \bibinfo{volume}{91} (\bibinfo{year}{2016}) \bibinfo{pages}{1--100}.

\bibitem[{Qin and Roberts(2020)}]{Qin:2020rad}
\bibinfo{author}{S.-X. Qin}, \bibinfo{author}{C.~D. Roberts},
  \bibinfo{title}{{Impressions of the Continuum Bound State Problem in QCD}},
  \bibinfo{journal}{Chin. Phys. Lett.}
  \bibinfo{volume}{37}~(\bibinfo{number}{12}) (\bibinfo{year}{2020})
  \bibinfo{pages}{121201}.

\bibitem[{Munczek(1995)}]{Munczek:1994zz}
\bibinfo{author}{H.~J. Munczek}, \bibinfo{title}{{Dynamical chiral symmetry
  breaking, Goldstone's theorem and the consistency of the Schwinger-Dyson and
  Bethe-Salpeter Equations}}, \bibinfo{journal}{Phys. Rev. D}
  \bibinfo{volume}{52} (\bibinfo{year}{1995}) \bibinfo{pages}{4736--4740}.

\bibitem[{Bender et~al.(1996)Bender, Roberts, and von Smekal}]{Bender:1996bb}
\bibinfo{author}{A.~Bender}, \bibinfo{author}{C.~D. Roberts},
  \bibinfo{author}{L.~von Smekal}, \bibinfo{title}{{Goldstone Theorem and
  Diquark Confinement Beyond Rainbow- Ladder Approximation}},
  \bibinfo{journal}{Phys. Lett. B} \bibinfo{volume}{380} (\bibinfo{year}{1996})
  \bibinfo{pages}{7--12}.

\bibitem[{Ding et~al.(2019)Ding, Raya, Bashir, Binosi, Chang, Chen, and
  Roberts}]{Ding:2018xwy}
\bibinfo{author}{M.~Ding}, \bibinfo{author}{K.~Raya},
  \bibinfo{author}{A.~Bashir}, \bibinfo{author}{D.~Binosi},
  \bibinfo{author}{L.~Chang}, \bibinfo{author}{M.~Chen}, \bibinfo{author}{C.~D.
  Roberts}, \bibinfo{title}{{$\gamma^\ast \gamma \to \eta, \eta^\prime$
  transition form factors}}, \bibinfo{journal}{Phys. Rev. D}
  \bibinfo{volume}{99} (\bibinfo{year}{2019}) \bibinfo{pages}{014014}.

\bibitem[{Binosi et~al.(2019)Binosi, Chang, Ding, Gao, Papavassiliou, and
  Roberts}]{Binosi:2018rht}
\bibinfo{author}{D.~Binosi}, \bibinfo{author}{L.~Chang},
  \bibinfo{author}{M.~Ding}, \bibinfo{author}{F.~Gao},
  \bibinfo{author}{J.~Papavassiliou}, \bibinfo{author}{C.~D. Roberts},
  \bibinfo{title}{{Distribution Amplitudes of Heavy-Light Mesons}},
  \bibinfo{journal}{Phys. Lett. B} \bibinfo{volume}{790} (\bibinfo{year}{2019})
  \bibinfo{pages}{257--262}.

\bibitem[{Wang et~al.(2018)Wang, Qin, Roberts, and Schmidt}]{Wang:2018kto}
\bibinfo{author}{Q.-W. Wang}, \bibinfo{author}{S.-X. Qin},
  \bibinfo{author}{C.~D. Roberts}, \bibinfo{author}{S.~M. Schmidt},
  \bibinfo{title}{{Proton tensor charges from a Poincar{\'e}-covariant Faddeev
  equation}}, \bibinfo{journal}{Phys. Rev. D} \bibinfo{volume}{98}
  (\bibinfo{year}{2018}) \bibinfo{pages}{054019}.

\bibitem[{Qin et~al.(2019)Qin, Roberts, and Schmidt}]{Qin:2019hgk}
\bibinfo{author}{S.-X. Qin}, \bibinfo{author}{C.~D. Roberts},
  \bibinfo{author}{S.~M. Schmidt}, \bibinfo{title}{{Spectrum of light- and
  heavy-baryons}}, \bibinfo{journal}{Few Body Syst.} \bibinfo{volume}{60}
  (\bibinfo{year}{2019}) \bibinfo{pages}{26}.

\bibitem[{Yao et~al.(2022)Yao, Binosi, Cui, and Roberts}]{Yao:2021pdy}
\bibinfo{author}{Z.-Q. Yao}, \bibinfo{author}{D.~Binosi},
  \bibinfo{author}{Z.-F. Cui}, \bibinfo{author}{C.~D. Roberts},
  \bibinfo{title}{{Semileptonic transitions: $B_{(s)} \to \pi(K)$; $D_s \to K$;
  $D\to \pi, K$; and $K\to \pi$}}, \bibinfo{journal}{Phys. Lett. B}
  \bibinfo{volume}{824} (\bibinfo{year}{2022}) \bibinfo{pages}{136793}.

\bibitem[{Roberts(1996)}]{Roberts:1994hh}
\bibinfo{author}{C.~D. Roberts}, \bibinfo{title}{{Electromagnetic pion
  form-factor and neutral pion decay width}}, \bibinfo{journal}{Nucl. Phys. A}
  \bibinfo{volume}{605} (\bibinfo{year}{1996}) \bibinfo{pages}{475--495}.

\bibitem[{Mezrag(2022)}]{Mezrag:2022pqk}
\bibinfo{author}{C.~Mezrag}, \bibinfo{title}{{An Introductory Lecture on
  Generalised Parton Distributions}}, \bibinfo{journal}{Few Body Syst.}
  \bibinfo{volume}{63}~(\bibinfo{number}{3}) (\bibinfo{year}{2022})
  \bibinfo{pages}{62}.

\bibitem[{Polyakov and Weiss(1999)}]{Polyakov:1999gs}
\bibinfo{author}{M.~V. Polyakov}, \bibinfo{author}{C.~Weiss},
  \bibinfo{title}{{Skewed and double distributions in pion and nucleon}},
  \bibinfo{journal}{Phys. Rev. D} \bibinfo{volume}{60} (\bibinfo{year}{1999})
  \bibinfo{pages}{114017}.

\bibitem[{Mezrag et~al.(2015)Mezrag, Chang, Moutarde, Roberts,
  Rodr{\'i}guez-Quintero, Sabati{\'e}, and Schmidt}]{Mezrag:2014jka}
\bibinfo{author}{C.~Mezrag}, \bibinfo{author}{L.~Chang},
  \bibinfo{author}{H.~Moutarde}, \bibinfo{author}{C.~D. Roberts},
  \bibinfo{author}{J.~Rodr{\'i}guez-Quintero},
  \bibinfo{author}{F.~Sabati{\'e}}, \bibinfo{author}{S.~M. Schmidt},
  \bibinfo{title}{{Sketching the pion's valence-quark generalised parton
  distribution}}, \bibinfo{journal}{Phys. Lett. B} \bibinfo{volume}{741}
  (\bibinfo{year}{2015}) \bibinfo{pages}{190--196}.

\bibitem[{Ding et~al.(2020)Ding, Raya, Binosi, Chang, Roberts, and
  Schmidt}]{Ding:2019lwe}
\bibinfo{author}{M.~Ding}, \bibinfo{author}{K.~Raya},
  \bibinfo{author}{D.~Binosi}, \bibinfo{author}{L.~Chang},
  \bibinfo{author}{C.~D. Roberts}, \bibinfo{author}{S.~M. Schmidt},
  \bibinfo{title}{{Symmetry, symmetry breaking, and pion parton
  distributions}}, \bibinfo{journal}{Phys. Rev. D}
  \bibinfo{volume}{101}~(\bibinfo{number}{5}) (\bibinfo{year}{2020})
  \bibinfo{pages}{054014}.

\bibitem[{Maris and Tandy(2000{\natexlab{a}})}]{Maris:2000sk}
\bibinfo{author}{P.~Maris}, \bibinfo{author}{P.~C. Tandy}, \bibinfo{title}{{The
  $\pi$, $K^+$, and $K^0$ electromagnetic form factors}},
  \bibinfo{journal}{Phys. Rev. C} \bibinfo{volume}{62}
  (\bibinfo{year}{2000}{\natexlab{a}}) \bibinfo{pages}{055204}.

\bibitem[{Dokshitzer(1977)}]{Dokshitzer:1977sg}
\bibinfo{author}{Y.~L. Dokshitzer}, \bibinfo{title}{Calculation of the
  Structure Functions for Deep Inelastic Scattering and $e^+$ $e^-$
  Annihilation by Perturbation Theory in Quantum Chromodynamics. ({\mbox {I}n
  {R}ussian})}, \bibinfo{journal}{Sov. Phys. JETP} \bibinfo{volume}{46}
  (\bibinfo{year}{1977}) \bibinfo{pages}{641--653}.

\bibitem[{Gribov and Lipatov(1971)}]{Gribov:1971zn}
\bibinfo{author}{V.~N. Gribov}, \bibinfo{author}{L.~N. Lipatov},
  \bibinfo{title}{{Deep inelastic electron scattering in perturbation theory}},
  \bibinfo{journal}{Phys. Lett. B} \bibinfo{volume}{37} (\bibinfo{year}{1971})
  \bibinfo{pages}{78--80}.

\bibitem[{Lipatov(1975)}]{Lipatov:1974qm}
\bibinfo{author}{L.~N. Lipatov}, \bibinfo{title}{{The parton model and
  perturbation theory}}, \bibinfo{journal}{Sov. J. Nucl. Phys.}
  \bibinfo{volume}{20} (\bibinfo{year}{1975}) \bibinfo{pages}{94--102}.

\bibitem[{Altarelli and Parisi(1977)}]{Altarelli:1977zs}
\bibinfo{author}{G.~Altarelli}, \bibinfo{author}{G.~Parisi},
  \bibinfo{title}{{Asymptotic Freedom in Parton Language}},
  \bibinfo{journal}{Nucl. Phys. B} \bibinfo{volume}{126} (\bibinfo{year}{1977})
  \bibinfo{pages}{298--318}.

\bibitem[{Cui et~al.(2022{\natexlab{b}})Cui, Ding, Morgado, Raya, Binosi,
  Chang, De~Soto, Roberts, Rodr\'\i{}guez-Quintero, and Schmidt}]{Cui:2022bxn}
\bibinfo{author}{Z.~F. Cui}, \bibinfo{author}{M.~Ding}, \bibinfo{author}{J.~M.
  Morgado}, \bibinfo{author}{K.~Raya}, \bibinfo{author}{D.~Binosi},
  \bibinfo{author}{L.~Chang}, \bibinfo{author}{F.~De~Soto},
  \bibinfo{author}{C.~D. Roberts},
  \bibinfo{author}{J.~Rodr\'\i{}guez-Quintero}, \bibinfo{author}{S.~M.
  Schmidt}, \bibinfo{title}{{Emergence of pion parton distributions}},
  \bibinfo{journal}{Phys. Rev. D} \bibinfo{volume}{105}~(\bibinfo{number}{9})
  (\bibinfo{year}{2022}{\natexlab{b}}) \bibinfo{pages}{L091502}.

\bibitem[{Yin et~al.(2023)Yin, Xu, Cui, Roberts, and
  Rodr\'\i{}guez-Quintero}]{Yin:2023dbw}
\bibinfo{author}{P.-L. Yin}, \bibinfo{author}{Y.-Z. Xu}, \bibinfo{author}{Z.-F.
  Cui}, \bibinfo{author}{C.~D. Roberts},
  \bibinfo{author}{J.~Rodr\'\i{}guez-Quintero}, \bibinfo{title}{{All-Orders
  Evolution of Parton Distributions: Principle, Practice, and Predictions}},
  \bibinfo{journal}{Chin. Phys. Lett. \emph{Express}}
  \bibinfo{volume}{40}~(\bibinfo{number}{9}) (\bibinfo{year}{2023})
  \bibinfo{pages}{091201}.

\bibitem[{Raya et~al.(2016)Raya, Chang, Bashir, Cobos-Martinez,
  Guti{\'e}rrez-Guerrero, Roberts, and Tandy}]{Raya:2015gva}
\bibinfo{author}{K.~Raya}, \bibinfo{author}{L.~Chang},
  \bibinfo{author}{A.~Bashir}, \bibinfo{author}{J.~J. Cobos-Martinez},
  \bibinfo{author}{L.~X. Guti{\'e}rrez-Guerrero}, \bibinfo{author}{C.~D.
  Roberts}, \bibinfo{author}{P.~C. Tandy}, \bibinfo{title}{{Structure of the
  neutral pion and its electromagnetic transition form factor}},
  \bibinfo{journal}{Phys. Rev. D} \bibinfo{volume}{93} (\bibinfo{year}{2016})
  \bibinfo{pages}{074017}.

\bibitem[{Maris and Tandy(2000{\natexlab{b}})}]{Maris:1999bh}
\bibinfo{author}{P.~Maris}, \bibinfo{author}{P.~C. Tandy}, \bibinfo{title}{{The
  quark photon vertex and the pion charge radius}}, \bibinfo{journal}{Phys.
  Rev. C} \bibinfo{volume}{61} (\bibinfo{year}{2000}{\natexlab{b}})
  \bibinfo{pages}{045202}.

\bibitem[{Roberts and Schmidt(2000)}]{Roberts:2000aa}
\bibinfo{author}{C.~D. Roberts}, \bibinfo{author}{S.~M. Schmidt},
  \bibinfo{title}{{Dyson-Schwinger equations: Density, temperature and
  continuum strong QCD}}, \bibinfo{journal}{Prog. Part. Nucl. Phys.}
  \bibinfo{volume}{45} (\bibinfo{year}{2000}) \bibinfo{pages}{S1--S103}.

\bibitem[{Xu et~al.(2021)Xu, Chen, Yao, Binosi, Cui, and Roberts}]{Xu:2021mju}
\bibinfo{author}{Y.-Z. Xu}, \bibinfo{author}{S.~Chen}, \bibinfo{author}{Z.-Q.
  Yao}, \bibinfo{author}{D.~Binosi}, \bibinfo{author}{Z.-F. Cui},
  \bibinfo{author}{C.~D. Roberts}, \bibinfo{title}{{Vector-meson production and
  vector meson dominance}}, \bibinfo{journal}{Eur. Phys. J. C}
  \bibinfo{volume}{81} (\bibinfo{year}{2021}) \bibinfo{pages}{895}.

\bibitem[{Ball and Chiu(1980)}]{Ball:1980ay}
\bibinfo{author}{J.~S. Ball}, \bibinfo{author}{T.-W. Chiu},
  \bibinfo{title}{{Analytic Properties of the Vertex Function in Gauge
  Theories. 1}}, \bibinfo{journal}{Phys.\ Rev.\ D} \bibinfo{volume}{22}
  (\bibinfo{year}{1980}) \bibinfo{pages}{2542--2549}.

\bibitem[{Curtis and Pennington(1990)}]{Curtis:1990zs}
\bibinfo{author}{D.~C. Curtis}, \bibinfo{author}{M.~R. Pennington},
  \bibinfo{title}{{Truncating the Schwinger-Dyson equations: How multiplicative
  renormalizability and the Ward identity restrict the three point vertex in
  QED}}, \bibinfo{journal}{Phys.\ Rev.\ D} \bibinfo{volume}{42}
  (\bibinfo{year}{1990}) \bibinfo{pages}{4165--4169}.

\bibitem[{Qin et~al.(2013)Qin, Chang, Liu, Roberts, and Schmidt}]{Qin:2013mta}
\bibinfo{author}{S.-X. Qin}, \bibinfo{author}{L.~Chang}, \bibinfo{author}{Y.-X.
  Liu}, \bibinfo{author}{C.~D. Roberts}, \bibinfo{author}{S.~M. Schmidt},
  \bibinfo{title}{{Practical corollaries of transverse Ward-Green-Takahashi
  identities}}, \bibinfo{journal}{Phys.\ Lett.\ B} \bibinfo{volume}{722}
  (\bibinfo{year}{2013}) \bibinfo{pages}{384--388}.

\bibitem[{Brout and Englert(1966)}]{Brout:1966oea}
\bibinfo{author}{R.~Brout}, \bibinfo{author}{F.~Englert},
  \bibinfo{title}{{Gravitational Ward Identity and the Principle of
  Equivalence}}, \bibinfo{journal}{Phys. Rev.}
  \bibinfo{volume}{141}~(\bibinfo{number}{4}) (\bibinfo{year}{1966})
  \bibinfo{pages}{1231--1232}.

\bibitem[{Raman(1971)}]{Raman:1971jg}
\bibinfo{author}{K.~Raman}, \bibinfo{title}{{Gravitational form-factors of
  pseudoscalar mesons, stress-tensor-current commutation relations, and
  deviations from tensor- and scalar-meson dominance}}, \bibinfo{journal}{Phys.
  Rev. D} \bibinfo{volume}{4} (\bibinfo{year}{1971}) \bibinfo{pages}{476--488}.

\bibitem[{Theussl et~al.(2004)Theussl, Noguera, and Vento}]{Theussl:2002xp}
\bibinfo{author}{L.~Theussl}, \bibinfo{author}{S.~Noguera},
  \bibinfo{author}{V.~Vento}, \bibinfo{title}{{Generalized parton distributions
  of the pion in a Bethe-Salpeter approach}}, \bibinfo{journal}{Eur. Phys. J.
  A} \bibinfo{volume}{20} (\bibinfo{year}{2004}) \bibinfo{pages}{483--498}.

\bibitem[{Xing et~al.(2023{\natexlab{c}})Xing, Ding, and Chang}]{Xing:2022mvk}
\bibinfo{author}{Z.~Xing}, \bibinfo{author}{M.~Ding},
  \bibinfo{author}{L.~Chang}, \bibinfo{title}{{Glimpse into the pion
  gravitational form factor}}, \bibinfo{journal}{Phys. Rev. D}
  \bibinfo{volume}{107}~(\bibinfo{number}{3})
  (\bibinfo{year}{2023}{\natexlab{c}}) \bibinfo{pages}{L031502}.

\bibitem[{Qin et~al.(2011)Qin, Chang, Liu, Roberts, and Wilson}]{Qin:2011dd}
\bibinfo{author}{S.-X. Qin}, \bibinfo{author}{L.~Chang}, \bibinfo{author}{Y.-X.
  Liu}, \bibinfo{author}{C.~D. Roberts}, \bibinfo{author}{D.~J. Wilson},
  \bibinfo{title}{{Interaction model for the gap equation}},
  \bibinfo{journal}{Phys. Rev. C} \bibinfo{volume}{84} (\bibinfo{year}{2011})
  \bibinfo{pages}{042202(R)}.

\bibitem[{Binosi et~al.(2015)Binosi, Chang, Papavassiliou, and
  Roberts}]{Binosi:2014aea}
\bibinfo{author}{D.~Binosi}, \bibinfo{author}{L.~Chang},
  \bibinfo{author}{J.~Papavassiliou}, \bibinfo{author}{C.~D. Roberts},
  \bibinfo{title}{{Bridging a gap between continuum-QCD and \emph{ab initio}
  predictions of hadron observables}}, \bibinfo{journal}{Phys. Lett. B}
  \bibinfo{volume}{742} (\bibinfo{year}{2015}) \bibinfo{pages}{183--188}.

\bibitem[{Chang et~al.(2009)Chang, Liu, Roberts, Shi, Sun, and
  Zong}]{Chang:2008ec}
\bibinfo{author}{L.~Chang}, \bibinfo{author}{Y.-X. Liu}, \bibinfo{author}{C.~D.
  Roberts}, \bibinfo{author}{Y.-M. Shi}, \bibinfo{author}{W.-M. Sun},
  \bibinfo{author}{H.-S. Zong}, \bibinfo{title}{{Chiral susceptibility and the
  scalar Ward identity}}, \bibinfo{journal}{Phys. Rev. C} \bibinfo{volume}{79}
  (\bibinfo{year}{2009}) \bibinfo{pages}{035209}.

\bibitem[{Xu et~al.(2023{\natexlab{b}})Xu, Yao, Qin, Cui, and
  Roberts}]{Xu:2022kng}
\bibinfo{author}{Z.-N. Xu}, \bibinfo{author}{Z.-Q. Yao}, \bibinfo{author}{S.-X.
  Qin}, \bibinfo{author}{Z.-F. Cui}, \bibinfo{author}{C.~D. Roberts},
  \bibinfo{title}{{Bethe\textendash{}Salpeter kernel and properties of
  strange-quark mesons}}, \bibinfo{journal}{Eur. Phys. J. A}
  \bibinfo{volume}{59}~(\bibinfo{number}{3})
  (\bibinfo{year}{2023}{\natexlab{b}}) \bibinfo{pages}{39}.

\bibitem[{Workman et~al.(2022)}]{Workman:2022ynf}
\bibinfo{author}{R.~L. Workman}, et~al., \bibinfo{title}{{Review of Particle
  Physics}}, \bibinfo{journal}{PTEP} \bibinfo{volume}{2022}
  (\bibinfo{year}{2022}) \bibinfo{pages}{083C01}.

\bibitem[{Maris and Roberts(1997)}]{Maris:1997tm}
\bibinfo{author}{P.~Maris}, \bibinfo{author}{C.~D. Roberts},
  \bibinfo{title}{{{$\pi$} and {$K$} meson Bethe-Salpeter amplitudes}},
  \bibinfo{journal}{Phys. Rev. C} \bibinfo{volume}{56} (\bibinfo{year}{1997})
  \bibinfo{pages}{3369--3383}.

\bibitem[{Krassnigg(2008)}]{Krassnigg:2009gd}
\bibinfo{author}{A.~Krassnigg}, \bibinfo{title}{{Excited mesons in a
  Bethe-Salpeter approach}}, \bibinfo{journal}{PoS}
  \bibinfo{volume}{CONFINEMENT\,8} (\bibinfo{year}{2008}) \bibinfo{pages}{075}.

\bibitem[{Krassnigg(2009)}]{Krassnigg:2009zh}
\bibinfo{author}{A.~Krassnigg}, \bibinfo{title}{{Survey of J=0,1 mesons in a
  Bethe-Salpeter approach}}, \bibinfo{journal}{Phys. Rev. D}
  \bibinfo{volume}{80} (\bibinfo{year}{2009}) \bibinfo{pages}{114010}.

\bibitem[{Llewellyn-Smith(1969)}]{LlewellynSmith:1969az}
\bibinfo{author}{C.~H. Llewellyn-Smith}, \bibinfo{title}{{A relativistic
  formulation for the quark model for mesons}}, \bibinfo{journal}{Annals Phys.}
  \bibinfo{volume}{53} (\bibinfo{year}{1969}) \bibinfo{pages}{521--558}.

\bibitem[{Polyakov and Schweitzer(2018)}]{Polyakov:2018zvc}
\bibinfo{author}{M.~V. Polyakov}, \bibinfo{author}{P.~Schweitzer},
  \bibinfo{title}{{Forces inside hadrons: pressure, surface tension, mechanical
  radius, and all that}}, \bibinfo{journal}{Int. J. Mod. Phys. A}
  \bibinfo{volume}{33}~(\bibinfo{number}{26}) (\bibinfo{year}{2018})
  \bibinfo{pages}{1830025}.

\bibitem[{Nakanishi(1969)}]{Nakanishi:1969ph}
\bibinfo{author}{N.~Nakanishi}, \bibinfo{title}{{A General survey of the theory
  of the Bethe-Salpeter equation}}, \bibinfo{journal}{Prog. Theor. Phys.
  Suppl.} \bibinfo{volume}{43} (\bibinfo{year}{1969}) \bibinfo{pages}{1--81}.

\bibitem[{Cui et~al.(2021)Cui, Binosi, Roberts, and Schmidt}]{Cui:2021aee}
\bibinfo{author}{Z.-F. Cui}, \bibinfo{author}{D.~Binosi},
  \bibinfo{author}{C.~D. Roberts}, \bibinfo{author}{S.~M. Schmidt},
  \bibinfo{title}{{Pion charge radius from pion+electron elastic scattering
  data}}, \bibinfo{journal}{Phys. Lett. B} \bibinfo{volume}{822}
  (\bibinfo{year}{2021}) \bibinfo{pages}{136631}.

\bibitem[{Chouika et~al.(2018)Chouika, Mezrag, Moutarde, and
  Rodr{\'{\i}}guez-Quintero}]{Chouika:2017rzs}
\bibinfo{author}{N.~Chouika}, \bibinfo{author}{C.~Mezrag},
  \bibinfo{author}{H.~Moutarde},
  \bibinfo{author}{J.~Rodr{\'{\i}}guez-Quintero}, \bibinfo{title}{{A
  Nakanishi-based model illustrating the covariant extension of the pion GPD
  overlap representation and its ambiguities}}, \bibinfo{journal}{Phys. Lett.
  B} \bibinfo{volume}{780} (\bibinfo{year}{2018}) \bibinfo{pages}{287--293}.

\bibitem[{Kumano et~al.(2018)Kumano, Song, and Teryaev}]{Kumano:2017lhr}
\bibinfo{author}{S.~Kumano}, \bibinfo{author}{Q.-T. Song},
  \bibinfo{author}{O.~V. Teryaev}, \bibinfo{title}{{Hadron tomography by
  generalized distribution amplitudes in pion-pair production process $\gamma^*
  \gamma \rightarrow \pi^0 \pi^0 $ and gravitational form factors for pion}},
  \bibinfo{journal}{Phys. Rev. D} \bibinfo{volume}{97} (\bibinfo{year}{2018})
  \bibinfo{pages}{014020}.

\bibitem[{Badier et~al.(1980)}]{Badier:1980jq}
\bibinfo{author}{J.~Badier}, et~al., \bibinfo{title}{{Measurement of the {$K^-
  / \pi^-$} structure function ratio using the Drell-Yan process}},
  \bibinfo{journal}{Phys. Lett. B} \bibinfo{volume}{93} (\bibinfo{year}{1980})
  \bibinfo{pages}{354}.

\bibitem[{Lin et~al.(2021)Lin, Chen, Fan, Zhang, and Zhang}]{Lin:2020ssv}
\bibinfo{author}{H.-W. Lin}, \bibinfo{author}{J.-W. Chen},
  \bibinfo{author}{Z.~Fan}, \bibinfo{author}{J.-H. Zhang},
  \bibinfo{author}{R.~Zhang}, \bibinfo{title}{{Valence-Quark Distribution of
  the Kaon and Pion from Lattice QCD}}, \bibinfo{journal}{Phys. Rev. D}
  \bibinfo{volume}{103}~(\bibinfo{number}{1}) (\bibinfo{year}{2021})
  \bibinfo{pages}{014516}.

\bibitem[{Shi et~al.(2014)Shi, Chang, Roberts, Schmidt, Tandy, and
  Zong}]{Shi:2014uwa}
\bibinfo{author}{C.~Shi}, \bibinfo{author}{L.~Chang}, \bibinfo{author}{C.~D.
  Roberts}, \bibinfo{author}{S.~M. Schmidt}, \bibinfo{author}{P.~C. Tandy},
  \bibinfo{author}{H.-S. Zong}, \bibinfo{title}{{Flavour symmetry breaking in
  the kaon parton distribution amplitude}}, \bibinfo{journal}{Phys. Lett. B}
  \bibinfo{volume}{738} (\bibinfo{year}{2014}) \bibinfo{pages}{512--518}.

\bibitem[{Binosi et~al.(2017)Binosi, Chang, Papavassiliou, Qin, and
  Roberts}]{Binosi:2016wcx}
\bibinfo{author}{D.~Binosi}, \bibinfo{author}{L.~Chang},
  \bibinfo{author}{J.~Papavassiliou}, \bibinfo{author}{S.-X. Qin},
  \bibinfo{author}{C.~D. Roberts}, \bibinfo{title}{{Natural constraints on the
  gluon-quark vertex}}, \bibinfo{journal}{Phys. Rev. D} \bibinfo{volume}{95}
  (\bibinfo{year}{2017}) \bibinfo{pages}{031501(R)}.

\bibitem[{Freese and Miller(2023)}]{Freese:2022fat}
\bibinfo{author}{A.~Freese}, \bibinfo{author}{G.~A. Miller},
  \bibinfo{title}{{Convolution formalism for defining densities of hadrons}},
  \bibinfo{journal}{Phys. Rev. D} \bibinfo{volume}{108}~(\bibinfo{number}{3})
  (\bibinfo{year}{2023}) \bibinfo{pages}{034008}.

\bibitem[{Zhang et~al.(2021{\natexlab{b}})Zhang, Cui, Ping, and
  Roberts}]{Zhang:2020ecj}
\bibinfo{author}{J.-L. Zhang}, \bibinfo{author}{Z.-F. Cui},
  \bibinfo{author}{J.~Ping}, \bibinfo{author}{C.~D. Roberts},
  \bibinfo{title}{{Contact interaction analysis of pion GTMDs}},
  \bibinfo{journal}{Eur. Phys. J. C} \bibinfo{volume}{81}~(\bibinfo{number}{1})
  (\bibinfo{year}{2021}{\natexlab{b}}) \bibinfo{pages}{6}.

\bibitem[{{\"O}zel and Freire(2016)}]{Ozel:2016oaf}
\bibinfo{author}{F.~{\"O}zel}, \bibinfo{author}{P.~Freire},
  \bibinfo{title}{{Masses, Radii, and the Equation of State of Neutron Stars}},
  \bibinfo{journal}{Ann. Rev. Astron. Astrophys.} \bibinfo{volume}{54}
  (\bibinfo{year}{2016}) \bibinfo{pages}{401--440}.

\bibitem[{Barabanov et~al.(2021)}]{Barabanov:2020jvn}
\bibinfo{author}{M.~Y. Barabanov}, et~al., \bibinfo{title}{{Diquark
  Correlations in Hadron Physics: Origin, Impact and Evidence}},
  \bibinfo{journal}{Prog. Part. Nucl. Phys.} \bibinfo{volume}{116}
  (\bibinfo{year}{2021}) \bibinfo{pages}{103835}.

\bibitem[{Eichmann(2011)}]{Eichmann:2011vu}
\bibinfo{author}{G.~Eichmann}, \bibinfo{title}{{Nucleon electromagnetic form
  factors from the covariant Faddeev equation}}, \bibinfo{journal}{Phys. Rev.
  D} \bibinfo{volume}{84} (\bibinfo{year}{2011}) \bibinfo{pages}{014014}.

\end{thebibliography}

\end{document}